\documentclass[prd,a4paper,superscriptaddress,aps,showpacs,twocolumn,floats,nofootinbib]{revtex4}
\usepackage{epsfig}
\usepackage{graphicx}
\usepackage{amsmath}
\usepackage{amssymb}
\usepackage{color}
\usepackage{soul}
\usepackage{hyperref}
\usepackage{url}
\usepackage{float}

\newcommand{\be}{\begin{equation}}
\newcommand{\ee}{\end{equation}}

\newcommand{\bea}{\begin{eqnarray}}
\newcommand{\eea}{\end{eqnarray}}
\newcommand{\bean}{\begin{eqnarray*}}
\newcommand{\eean}{\end{eqnarray*}}

\newcommand{\LCDM}{\Lambda\rm{CDM}}

\begin{document}

\title{Non-minimal Derivative Coupling Scalar Field and Bulk Viscous Dark Energy}

\author{Behrang Mostaghel}
\email{b$_$mostaghel@sbu.ac.ir}

\affiliation{Department of Physics, Shahid Beheshti University,\\  G.C., Evin, Tehran 19839, Iran }

\author{Hossein Moshafi}
\email{hosseinmoshafi@iasbs.ac.ir}

\affiliation{Department of Physics, Institute for Advanced Studies in Basic Sciences, \\
Gavazang, Zanjan, IRAN}

\author{S.M.S. Movahed}
\email{m.s.movahed@ipm.ir}

\affiliation{Department of Physics, Shahid Beheshti University,\\  G.C., Evin, Tehran 19839, Iran }
\affiliation{School of Physics, Institute for Researches in Fundamental Sciences (IPM), P.O.Box 19395-5531,Tehran, Iran}

\date{\today}

\begin{abstract}

Inspired by thermodynamical dissipative phenomena, we consider bulk viscosity for dark fluid in a spatially flat two-component Universe. Our viscous dark energy model represents Phantom crossing avoiding Big-Rip singularity. We propose a non-minimal derivative coupling scalar field with zero potential leading to accelerated expansion of Universe in the framework of
bulk viscous dark energy model. In this approach, coupling constant ($\kappa$) is related to viscosity coefficient ($\gamma$) and energy density of dark energy at the present time ($\Omega_{\rm DE}^0$). This coupling is bounded as $\kappa\in [-1/9H_0^2(1-\Omega_{\rm DE}^0), 0]$ and  for $\gamma=0$ leads to $\kappa=0$.  To perform robust analysis, we implement recent observational data sets including Joint Light-curve Analysis (JLA) for SNIa, Gamma Ray Bursts (GRBs) for most luminous astrophysical objects at high redshifts, Baryon Acoustic Oscillations (BAO) from different surveys, Hubble parameter from HST project, {\it Planck} data for CMB power spectrum and CMB Lensing. Joint analysis of JLA$+$GRBs$+$BAO$+$HST shows that $\Omega_{\rm DE}^0=0.696\pm 0.010$, $\gamma=0.1404\pm0.0014$ and $H_0=68.1\pm1.3$ at $1\sigma$ confidence interval.  {\it Planck} TT observation provides $\gamma=0.32^{+0.31}_{-0.26}$ at $68\%$ confidence limit for viscosity coefficient. Tension in Hubble parameter is alleviated in this model. Cosmographic distance ratio indicates that current observed data prefer to increase bulk viscosity.  Finally, the competition between Phantom and Quintessence behavior of viscous dark energy model can accommodate cosmological old objects reported as a sign of age crisis in $\Lambda$CDM model.
\keywords{Dark Energy \and Bulk Viscous Cosmology \and Non-minimal derivative coupling scalar field}

\end{abstract}

\pacs{98.80.-k, 98.80.Es, 98.80.Jk, 98.65.Dx}

\maketitle

\vspace{0.1cm}
\section{Introduction}
Accelerating expansion of the Universe has widely been confirmed by increasing number of consequences of observational data \cite{Riess:1998cb,Perlmutter:1998np}.  Various observations such as Supernova Type Ia, Cosmic Microwave Background (CMB) and Baryonic Acoustic Oscillations (BAO) have indicated the break down of general relativity (GR) at cosmic scale  \cite{Ade:2015rim}. Many scenarios have been proposed to produce a repulsive force in order to elucidate the current accelerating epoch. 

There are three approaches beyond model including cosmological constant \cite{Copeland:2006wr,Amendola:2012ys,Nojiri:2010wj,Bull:2015stt}: first approach corresponds to dynamical dark energy incorporating field theoretical orientation and phenomenological dark fluids. Second category is devoted to modified general relativity including  Horndeski's types such as, Galileons, Chameleons, Brans-Dicke, Symmetrons, and other possibilities \cite{Amendola:2012ys,Nojiri:2010wj,Bull:2015stt,Horndeski:1974wa,Copeland:2012qf,Charmousis:2011bf,Appleby:2012rx,Konnig:2016idp}. In third strategy, based on thermodynamics point of view, phenomenological exotic fluids are supposed for alternative dark energy model \cite{Kamenshchik:2001cp,Fabris:2001tm,Bento:2002ps}.

Generally, proposing exotic fluids for various applications has historically been highlighted in the first order irreversible processes \cite{Eckart:1940te} and also in second order correction to the dissipative process proposed via  W. Israel \cite{Israel_1979}. Dissipative process is a generic property of any realistic phenomena. Accordingly, bulk and shear viscous terms are most relevant parts which should be taken into account for a feasible relativistic fluid \cite{Szyd_owski_2007,Avelino:2008ph,belinskii79,Brevik_2005,Cataldo_2005,Di_Prisco_2000}. In other words, to recognize the mechanism of heat generation and characteristic of smallest fluctuations one possibility is revealing the presence of dissipative processes \cite{Weinberg:1971mx}. In classical fluid dynamics, viscosity is trivial consequence of internal  degree of freedom of the associated particles \cite{van_den_Horn_2016}. The existence of any dissipative process for isotropic and homogeneous Universe, should undoubtedly be scalar, consequently, at the background level, we deal with bulk viscous model \cite{Szyd_owski_2007,Avelino:2008ph,Brevik_2005,Zimdahl:2000zm,Velten_2013,Sasidharan_2015,Capozziello_2006}. 
Some examples to describe the source of viscosity are as follows: moving cosmic strings through the cosmic magnetic fields, magnetic monopoles in monopole interactions effectively experience various viscous phenomena \cite{Ostriker_1986,Hu:1986jd}.  Various mechanisms for primordial quantum particle productions and their interactions were also main persuading for viscous fluid \cite{Weinberg:1971mx,Hu:1986jd,Barrow_1986,Gr_n_1990,Maartens_1995,Eshaghi:2015tqa}.

In the context of exotic fluids, a possible approach to explain the late time accelerating expansion of the Universe is constructing an exotic fluid including bulk viscous term. A considerable  part of previous studies have been devoted to one-component for dark sector \cite{Brevik_2005,Sasidharan_2015,Normann_2016,Brevik_2015,WANG_2014}. A motivation behind mentioned proposals are avoiding from dark degeneracy \cite{Kunz_2009}. Bulk viscosity in cosmological models has also resolved the so-called Big-rip problem \cite{Frampton:2011sp,Brevik_2011}.  Recently, Normann et al., tried to map the viscous radiation or matter to a Phantom dark energy model and they showed that Phantom dark energy can be misinterpreted due to existence of non-equilibrium  pressure causing viscosity in pressure of either in matter or radiation \cite{Normann_2016}.

 {\it A priori} approach for bulk viscous cosmology encouraged some authors to build robust mechanisms to describe a correspondence for the mentioned dissipative term. Indeed without any reasonable model for underlying dissipative processes, one can not express any thing about the presence of such fluid including exotic properties \cite{Avelino:2008ph,Zimdahl:2000zm,Zimdahl_2000,Wilson_2007,Mathews_2008,Ren_2006}.

In cosmology, inspired by inflationary paradigm, canonical scalar fields with minimal coupling to gravity have been introduced to explain the origin of extraordinary matters \cite{Guth_1981,Ratra:1987rm,Caldwell:1999ew,Singh:2003vx,Zlatev_1999}. 
 Models with non-canonical scalar fields with minimal coupling or non-minimal coupling are other phenomenological descriptions to construct dark energy component \cite{Cai:2009zp,Uzan:1999ch}.  Another alternative approach is
non-minimal derivative coupling which appears in various approaches
such as, Jordan-Brans-Dicke theory \cite{Brans_1961}, quantum
field theory \cite{Birrell_1982} and low energy-limit of the
superstrings \cite{Appelquist_1983,Holden_2000,Easson_2007}.
Following a research done by L. Amendola, many
models for non-minimal derivative coupling (NMDC) have been proposed
to investigate inflationary epoch and late time accelerating
expansion
\cite{Amendola:1993uh,Germani:2010gm,Capozziello:1999uwa,Capozziello:1999xt,Granda:2010hb}.
Concentrating on cosmological applications of NMDC, S.V. Sushkhov et
al. showed that for a specific value of coupling constant, it is
possible to construct a new exact cosmological solution without
considering a certain form of scalar field potential
\cite{Sushkov:2009hk,Saridakis:2010mf}. It has been demonstrated
that a proper action containing a Lagrangian with NMDC scalar field
for dynamical dark energy model enables to solve Phantom crossing
\cite{Perivolaropoulos:2005yv,Banijamali:2012hn,Gao:2010vr}. {In a paper by Granda et. al, non-minimal kinetic coupling in the framework of Chaplygin cosmology has been considered \cite{Granda:2011zy}.}

In the present paper, we try to propose a modified version for dark
energy model inspired by dissipative phenomena in fluids with
following advantages and novelties: considering special type of
viscosity satisfying isotropic
property of cosmos at the background level. In this paper, to make
more obvious concerning the knowledge of bulk viscosity, we will
rely on modified general relativity { to obtain corresponding scalar field giving rise to accelerated expansion of Universe in context of bulk viscous dark energy scenario.}
With this mechanism, we will show the correspondence between our
viscous dark energy model and the scalar-tensor theories in a
two-component dark sectors model in contrast to that of done in
\cite{Frampton:2011sp,Brevik_2011}.
 In addition we will demonstrate that our viscous dark energy model without any interaction between dark sectors, has Phantom crossing avoiding Big-Rip singularity. Considering two components for dark sides of Universe in our approach leads to no bouncing behavior for consistent viscosity coefficient. Generally there is no ambiguity in computation of cosmos' age. Observational consequences indicates to resolve tension in Hubble parameter.

 From observational points of view, ongoing and upcoming generation of ground-based and space-based surveys classified in various stages ranging from I to IV,  one can refer to background and perturbations of observables \cite{Bull:2015stt,Amendola:2012ys}. Subsequently, we will rely on the state-of-the-art observational data sets such as Supernova Type Ia (SNIa), Gamma Ray Bursts (GRBs), Baryonic Acoustic Oscillation (BAO) and CMB evolution based on background dynamics to examine the consistency of our model. 
Here we have incorporated the contribution of viscous dark energy in the dynamics of background.

The rest of this paper is organized as follows: In section II we introduce our viscous dark energy model as a candidate of dark energy. Background dynamics of the Universe will be explained in this section. We use
Lagrangian approach with a non-minimal derivative coupling scalar field in order to provide a theoretical model for clarifying the correspondence of viscous dark energy,  in section III.  Effect of our model on the geometrical parameters of the Universe, namely, comoving distance, Alcock-Paczynski, comoving volume element, cosmographic parameters will be examined in section IV. To distinguish between viscous dark energy and cosmological constant as a dark energy, we will use $Om$-diagnostic test in the mentioned section. Recent observational data and the posterior analysis will be explore in section V. Section VI is devoted to results and discussion concerning the consistency of viscous dark energy model with the observations, cosmographic distance ratio, Hubble parameter and cosmic age crisis. Summary and concluding remarks are given in section VII.


\section{Bulk-viscous cosmology}
\label{Bulk-viscous cosmology}

In this section, we explain a model for dark energy produce accelerating expansion in the history of cosmos evolution. To this end, we consider a bulk viscous model, correspondingly, energy-momentum tensor will be modified. We also propose a new solution for mentioned model to construct so-called dynamical dark energy.

\subsection{Background Dynamics in the Presence of Bulk Viscosity}

Dynamics of the Universe is determined by the Einstein's field equations:
\begin{eqnarray}
G_{\mu \nu }=8\pi G_{N} T_{\mu \nu},
\end{eqnarray}
where $ G_{\mu \nu}=R_{\mu \nu}-\frac{1}{2}R g_{\mu \nu }$ is Einstein's tensor and $G_{N}$ is Newton's gravitational constant.  $T_{\mu \nu} $ is \textit{energy-momentum tensor} given by:
\begin{eqnarray}
T_{\mu \nu}=T_{\mu \nu}^{\rm{m}}+T_{\mu \nu}^{\rm{rad}}+ \overline{T}_{\mu \nu}.
\end{eqnarray}
where $ T^{\rm{m}}_{\mu \nu}$ and $T_{\mu \nu}^{\rm{rad}}$ are the energy-momentum tensor of the matter and radiation respectively. Generally, tensor $\overline{T}_{\mu \nu}$ includes other sources of gravity, such as scalar fields \cite{Caldwell:1999ew}. For cosmological constant, $\Lambda $, we define the energy-momentum tensor in the form of $\overline{T}_{\mu \nu}=-\frac{\Lambda}{8\pi G_{N}}g_{\mu \nu} $. Here we consider the following form for $T_{\mu \nu}$:
\begin{eqnarray}\label{perfect fluid energy momentum tensor}
T_{\mu \nu}=\left(\rho + p \right)u_{\mu} u_{\nu}+ pg_{\mu \nu}
+2q_{( \mu }u_{\nu )}+\pi_{\mu \nu},
\end{eqnarray}
in this equation, $\rho$ is energy density and $p$ is the pressure of the fluid,
$q_{\mu}=-\left( \delta^{\nu}_{\mu}+u_{\mu} u^{\nu}  \right)T_{\nu \alpha}u^{\alpha} $ is the energy flux vector, and $\pi_{\mu \nu} $ is the symmetric and traceless anisotropic stress tensor \cite{Tsagas:2007yx}. For \textit{barotropic fluid}, namely $p=p\left(\rho\right) $ case, $q_{\mu}$ and $\pi_{\mu \nu }$ are identically zero and one can define \textit{equation of state} (EoS) in the form of $w\left(\rho\right)=p\left(\rho\right)/\rho $. 
Applying the FLRW metric to the Einstein's equations with a given $T_{\mu\nu}$, gives Friedmann equations as follows:
\begin{eqnarray}
H^{2}+\frac{k}{a^{2}}&=&\frac{8\pi G_{N}}{3} \rho,\\ \nonumber
\frac{\ddot{a}}{a}&=&-\frac{4 \pi G_{N}}{3} \left(    \rho+ 3 p       \right),
\end{eqnarray}
where $\rho=\sum_{i}\rho_{i} $ , $p=\sum_{i}p_{i} $ and $H=\frac{\dot{a}}{a} $ are total energy density, total pressure and \textit{Hubble parameter}, respectively. Also $k$ indicates the geometry of the Universe. The continuity equation for all components read as:
\begin{eqnarray}
\frac{d\rho}{d t}+3 H \left(  \rho + p          \right)=0.
\end{eqnarray}
It turns out that, if there is no interaction between different components of the Universe, consequently, continuity equation for each component becomes:
\begin{eqnarray}\label{continuity eq}
\frac{d\rho_{i}}{dt}+3H \left(  \rho_{i}+p_{i}          \right)=0.
\end{eqnarray}
Now, it is possible to derive EoS parameter in general case, by solving following equation:
\begin{eqnarray}
w_{i}\equiv\frac{p_i}{\rho_i}=-1-\frac{1}{3} \frac{d \ln{\rho_{i}}}{d \ln{a}}.
\end{eqnarray}
In the next subsection, we will build a dark energy model, according to viscosity assumptions.

\begin{figure}[H]
\centering
\includegraphics[width=0.4\textwidth]{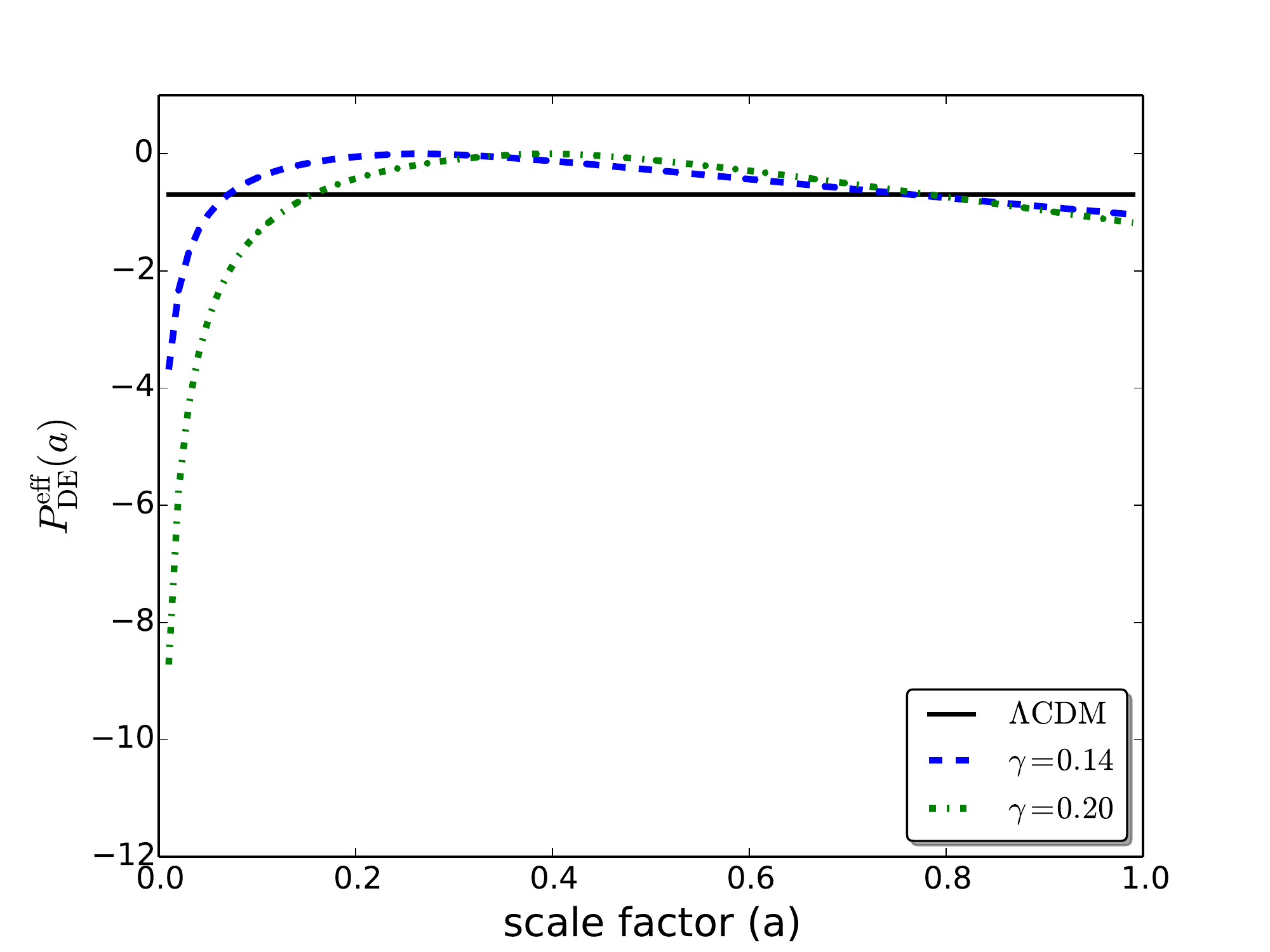}\\
\includegraphics[width=0.4\textwidth]{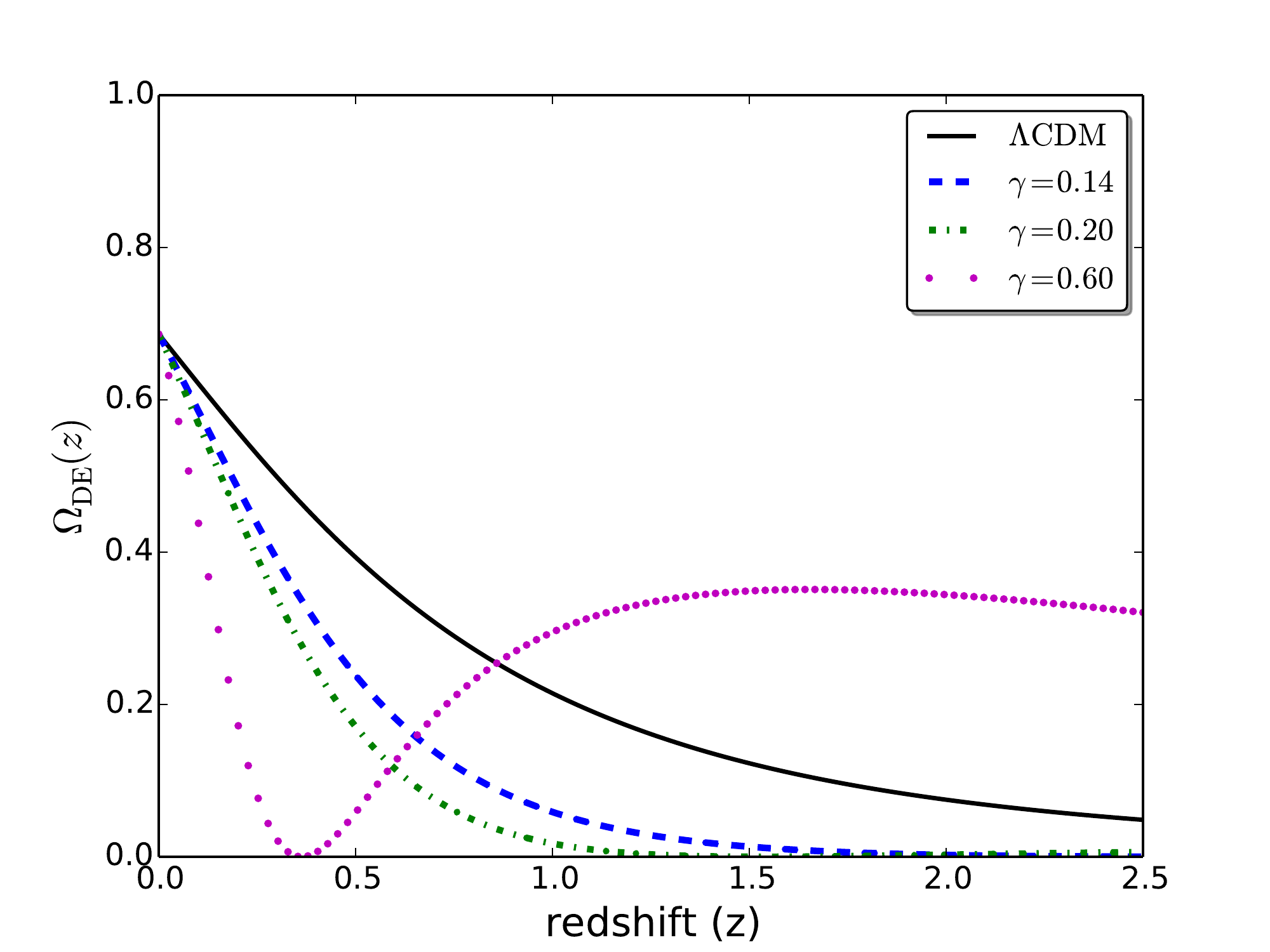}
     \caption{\textit{Upper panel:} Pressure of viscous fluid as a function of scale factor. \textit{Lower panel:} $\Omega_{\rm DE}(z)$ as a function of redshift. In both cases we changed the value of $\gamma$. The other free parameters have been fixed by SNIa constraint. $\Lambda$CDM best fit is given by {\it Planck} observation. }
\label{fig:pressure-a}
\end{figure}

\subsection{Viscous dark energy Model}
\label{proposed model}

In this subsection, we consider a bulk viscous fluid as a representative of so-called dark energy which is responsible of late time acceleration.
For a typical dissipative fluid, according to Ekart's theory as a first order limit of Israel-Stewart model with zero relation time, one can rewrite effective pressure, $p^{\rm{eff}}$, in the following form \cite{Eckart:1940te}:

\begin{eqnarray}
p^{\rm{eq}}\rightarrow p^{\rm{eff}}&=&p^{\rm{eq}}-\zeta \Theta(t)\\ \nonumber
&=&w\left(\rho\right)  \rho-\zeta \Theta(t),
\end{eqnarray}
where $p^{\rm{eq}}$ is pressure at thermodynamical equilibrium. $\zeta$ and $\Theta\left(t\right) $ are viscosity and expansion scalar, respectively. 
In general case, $\zeta$ is not constant and  there are many approaches to determine the functionality form of viscosity. In general case viscosity is a function for thermodynamical state, i.e, energy density of the fluid, $\zeta\left(\rho  \right)$ \cite{Li:2009mf,Hiscock:1991sp}. According to mentioned dissipative approach, we propose the
following model for the pressure of the viscous dark energy:
\begin{eqnarray}\label{ef-p}
p^{\rm{eff}}_{\rm DE}=-\rho_{\rm DE}-\zeta \Theta(t),
\end{eqnarray}
for FLRW cosmology, we have $\Theta(t)=3H(t)$. In the model that we consider throughout this paper, $\rho_{\rm DE}$ is dynamical variable due to its viscosity.  In principle, higher order corrections can be implemented in modified energy-momentum tensor, but it was demonstrated that mentioned terms have no considerable influence on the cosmic acceleration \cite{Hiscock:1991sp}. Therefore those relevant terms having dominant contributions for isotropic and homogeneous Universe at large scale phenomena are survived.
One can expect that the viscosity is affected by individual nature of corresponding energy density and implicitly is generally manipulated by expansion rate of the Universe,  accordingly, our ansatz about the dark energy viscosity is:
\begin{eqnarray}\label{ansatz}
\zeta(\rho_{\rm DE},H)=\xi\frac{\sqrt{\rho_{\rm DE}}}{H},
\end{eqnarray}
where coefficient $\xi$ is a positive constant and $H=\dot{a}/a$ is Hubble parameter.  According to  this choice, in the early Universe when the dark matter has dominant contribution epoch, viscosity of dark energy becomes negligible. 
On the other hand, at late time, this term increases and in dark energy dominated Universe leads to constant $\xi$. 

Inserting Eq. (\ref{ef-p}) in Eq. (\ref{continuity eq}) and using  Eq.(\ref{ansatz}) in the flat Universe for $\rho_{\rm DE}$ resulting  in
 the evolution of bulk viscous model for dark energy in terms of scale factor as:

\begin{eqnarray}\label{e-d}
\rho_{\rm DE}\left(a\right) = \rho_{\rm DE}^{0}
\left(1 +\frac{9 \xi}{2\sqrt{\rho^{0}_{\rm DE}}}\ln{a}            \right)^{2}.
\end{eqnarray}
\begin{figure}
\centering
    \includegraphics[width=.9\columnwidth]{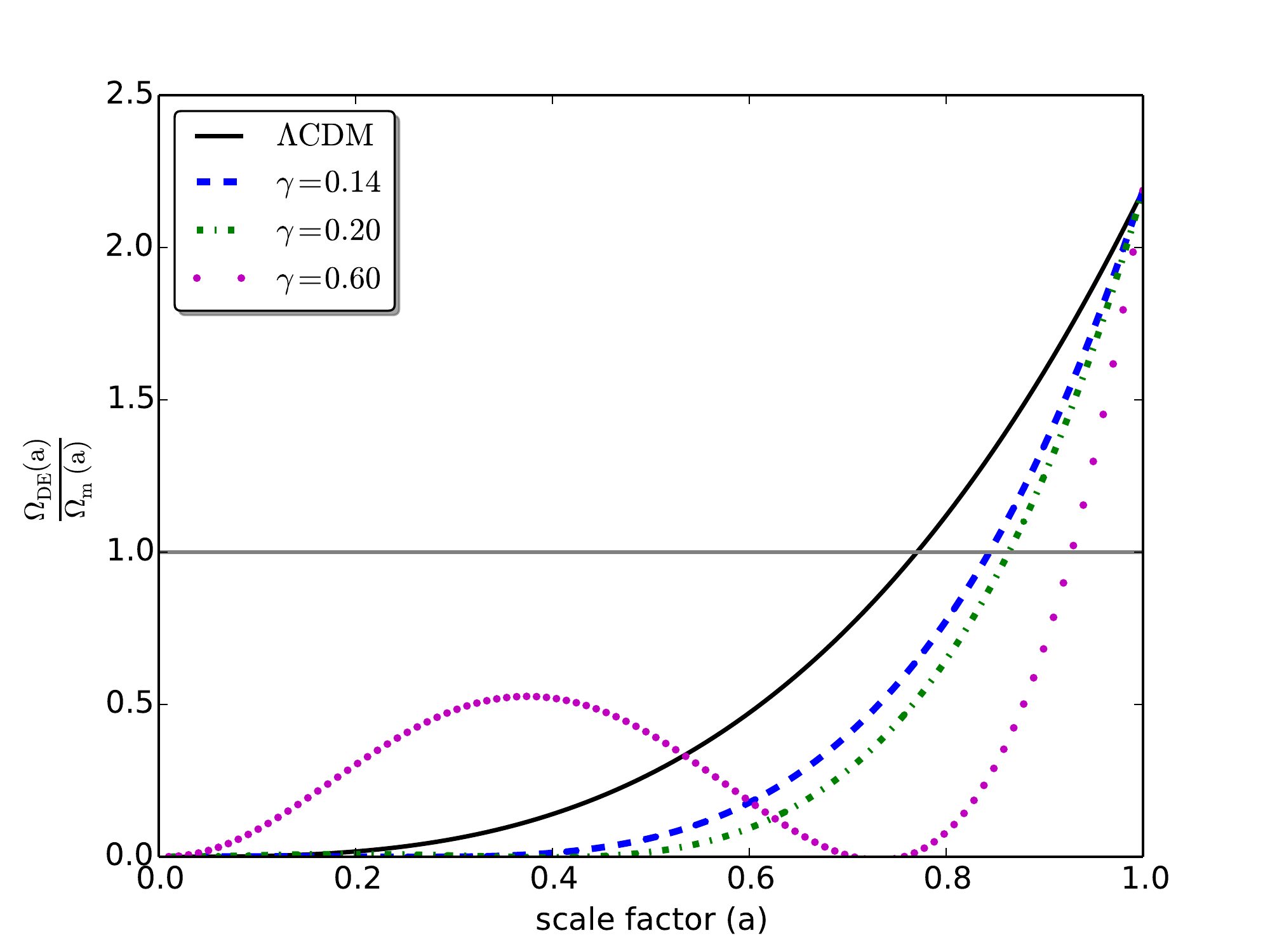}
    \caption{ Ratio of $\Omega_{\rm DE} (a)$ to $\Omega_m(a)$ as a function of scale factor. The other free parameters have been fixed by SNIa observational constraint.  $\Lambda$CDM best fit is given by {\it Planck} observation.}
\label{fig:density-ratio-a}
\end{figure}

\begin{widetext}
\begin{center}
\begin{table}[H]
\centering
\begin{tabular}{|c|c|c|c|c|c|}
\hline
\hline Type & Time & scale factor & Energy density& Pressure & EoS \\
\hline
I (Big-Rip)&$t\to t_s$&$a\to\infty$&$\rho\to \infty$&$|p|\to \infty$&undefined \\\hline
II (Sudden)&$t\to t_s$&$a\to a_s$&$\rho\to \rho_s$&$|p|\to \infty$&undefined \\\hline
III &$t\to t_s$&$a\to a_s$&$\rho\to \infty$&$|p|\to \infty$&undefined \\\hline
IV &$t\to t_s$&$a\to a_s$&$\rho\to 0$&$|p|\to 0$&undefined \\\hline
V (Little-Rip)&$t\to \infty$&$a\to\infty$&$\rho\to \infty$&$|p|\to \infty$& $w=p/\rho\to -1$ \\\hline
\hline
\end{tabular}
\caption{\label{tab:singularity} Various type of singularities in energy density and pressure given from  \cite{Nojiri:2005sx,Nojiri:2005sr,Stefancic:2004kb,Frampton:2011sp}. }
\end{table}
\end{center}
\end{widetext}

Dimensionless dark energy density can be written as:
\begin{eqnarray}\label{de-p}
\Omega_{\rm DE}(a)&=&\frac{H_{0}^{2}\Omega^{0}_{\rm DE}}{H^{2}} \left( 1+ \frac{9 \gamma}{2 \sqrt{\Omega^{0}_{\rm DE}}}  \ln{a}         \right)^{2}.
\end{eqnarray}
where $\Omega^{0}_{\rm DE}=8\pi G_N\rho_{\rm DE}^{0}/3H_0^2$ and $\gamma $ is  the dimensionless viscosity coefficient defined by:
\begin{eqnarray}\label{gamma}
\gamma\equiv \sqrt{\frac{8\pi G_{N}}{3H_{0}^2}}\xi.
\end{eqnarray}
Therefore the evolution of background in the present of viscous dark energy model reads as:
\begin{widetext}
\begin{eqnarray}\label{eq:hubble}
\left(\frac{\dot{a}}{a}\right)^2=H_0^2\left[\Omega_{\rm r}^0a^{-4}+\Omega_{\rm m}^0 a^{-3}+\Omega_{\rm DE}^0\left(1 +\frac{9 \gamma}{2\sqrt{\Omega^{0}_{\rm DE}}}\ln{a}\right)^{2}\right] +H_0^2(1-\Omega_{\rm tot}^0)a^{-2},
\end{eqnarray}
\end{widetext}
here $\Omega_{\rm tot}^0=\Omega_{\rm r}^0+\Omega_{\rm
m}^0+\Omega_{\rm DE}^0$ and throughout this paper we consider flat
Universe. In this paper we consider two component for dark sector of
Universe. Also we will show that in this case there is no bouncing
behavior presenting in one-component phenomenological fluid
considered in  ref. \cite{Frampton:2011sp}.

According to Eq. (\ref{de-p}), dark energy has a minimum at $\tilde{a}$, where:
\begin{eqnarray}\label{red-shift}
\tilde{a}=\exp{\left(\frac{-2 \sqrt{\Omega^{0}_{\rm DE}}}{9 \gamma}         \right)}.
\end{eqnarray}
and this value of minimum is equal to zero. 
The value of $\Omega_{\rm DE}(a)$ for $a\gg\tilde{a}$ and for $a\ll
\tilde{a}$ is  independent of the present value of dark energy,
$\left( \Omega_{\rm DE}^0\right)$, and this value is a pure effect
of bulk viscosity. To examine the variation of pressure and energy
density of viscous dark energy, we plot their behavior in Fig.
\ref{fig:pressure-a}. The upper panel of Fig. \ref{fig:pressure-a}
shows the effective pressure in this model while lower panel
corresponds to $\Omega_{\rm DE}(z)$. To make more sense, we compare
the behavior of this model with $\Lambda$CDM.
This figure demonstrates that at the late time, there is a
significant change in the behavior of viscous dark energy model,
consequently in order to distinguish between cosmological constant
and our model we should take into account those  indicators which
are more sensitive around late time. This kind of behavior probably
may affect on the structure formation of Universe and has unique
footprint on large scale structure.

The ratio of viscous dark energy to energy density of dark matter as
a function of scale factor indicates that the modified version of
dark energy has late time contribution in the expansion rate of
Universe (see Fig. \ref{fig:density-ratio-a}). It is worth noting
that for some values of viscosity, the relative contribution at early
epoch is higher than that of in $\Lambda$CDM while for the late time
this contribution is less than $\Lambda$CDM demonstrating an
almost oscillatory behavior.

Interestingly, according to Fig.~\ref{fig:pressure-a}, the viscous DE has non-monotonic evolution form decreasing interval to increasing situation depending on value of $\gamma$. This behavior this
demonstrates a crossing from Quintessence phase to the Phantom phase
corresponding to a Phantom crossing \cite{Brevik:2005bj,
Brevik:2006md}.
According to equation of state one can compute the effective form of  equation of state given by:
\begin{equation}
w_{\rm eff}(a)\equiv \frac{ \int_1^a w_{\rm DE}(a') d\ln a'}{\int_1^a d\ln a'}.
\end{equation}
By increasing the value of $\gamma$, we get the
so-called Phantom crossing behavior for viscous dark energy model at
late time. For positive (negative) value of viscosity coefficient, Universe at late time is dominated by Phantom (Quintessence) component. 
 It has been shown that for $w_{\rm DE}<-1$ cosmological models have future singularity.
According to previous studies, we summarize future singularities in cosmology in Table \ref{tab:singularity}. In mentioned table $t_s$ and $a_s$ are respectively characteristic time and scale factor, where divergence happens. For $a\to \infty$, the EoS of viscous dark energy goes asymptotically to $-1$ and consequently this model  belongs to \textit{Little-Rip} category \cite{Frampton:2011sp}. In addition, by using Eq. (\ref{eq:hubble}), we can write the age of cosmos at a given scale factor as: 
\begin{eqnarray}
t_0-t(a)=\int_a^1{\frac{d a}{H_0a \sqrt{\left(1-\Omega^{0}_{\rm DE} \right)a^{-3}+     \Omega^{0}_{\rm DE} \left( 1- \frac{\ln{a}}{\ln{\tilde{a}}}         \right)^{2}    }  }  }.
\end{eqnarray}
Solving scale factor as a function of time for dark energy dominated era leads to:
\begin{eqnarray}
\lim_{t\rightarrow \infty}a\left(t\right )\simeq 
\exp{\left( e^{\frac{9\gamma  H_{0}}{2}\left(t-t_{0}\right)}\right)}.
\end{eqnarray}
 therefore, energy density reads as:
\begin{eqnarray}
\lim_{t\rightarrow \infty}\rho_{\rm DE}\left(t\right) \simeq \rho^{0}_{\rm DE} e^{9\gamma H_{0}\left(t-t_{0}\right)} .
\end{eqnarray}
According to the above equation, at infinite time, the energy density of the dark energy reaches to infinity when $t\to \infty$, demonstrating our model has Little-Rip singularity. 
In one component Universe, the age of Universe can be determined via:
\begin{eqnarray}\label{ageb}
\int{d t}=\int{\frac{d a}{H_0\Omega^{0}_{\rm DE} a\left(1-\frac{\ln{a}}{\ln{\tilde{a}}}  \right) } }\\
\nonumber
=\frac{1}{H_0 \Omega^{0}_{\rm DE}}\frac{\ln{ \frac{\ln{a}}{\ln{\tilde{a}}}}}{\ln{(1/\tilde{a})}}.
\end{eqnarray}
Subsequently, for $a=\tilde{a}$,  time is undefined which is a property of bouncing model \cite{Novello:2008ra}. In order to avoid mentioned case, we consider two component Universe in spite of that of considered in ref. \cite {Frampton:2011sp}.

\begin{figure}[H]
\centering
  \includegraphics[width=.9\columnwidth]{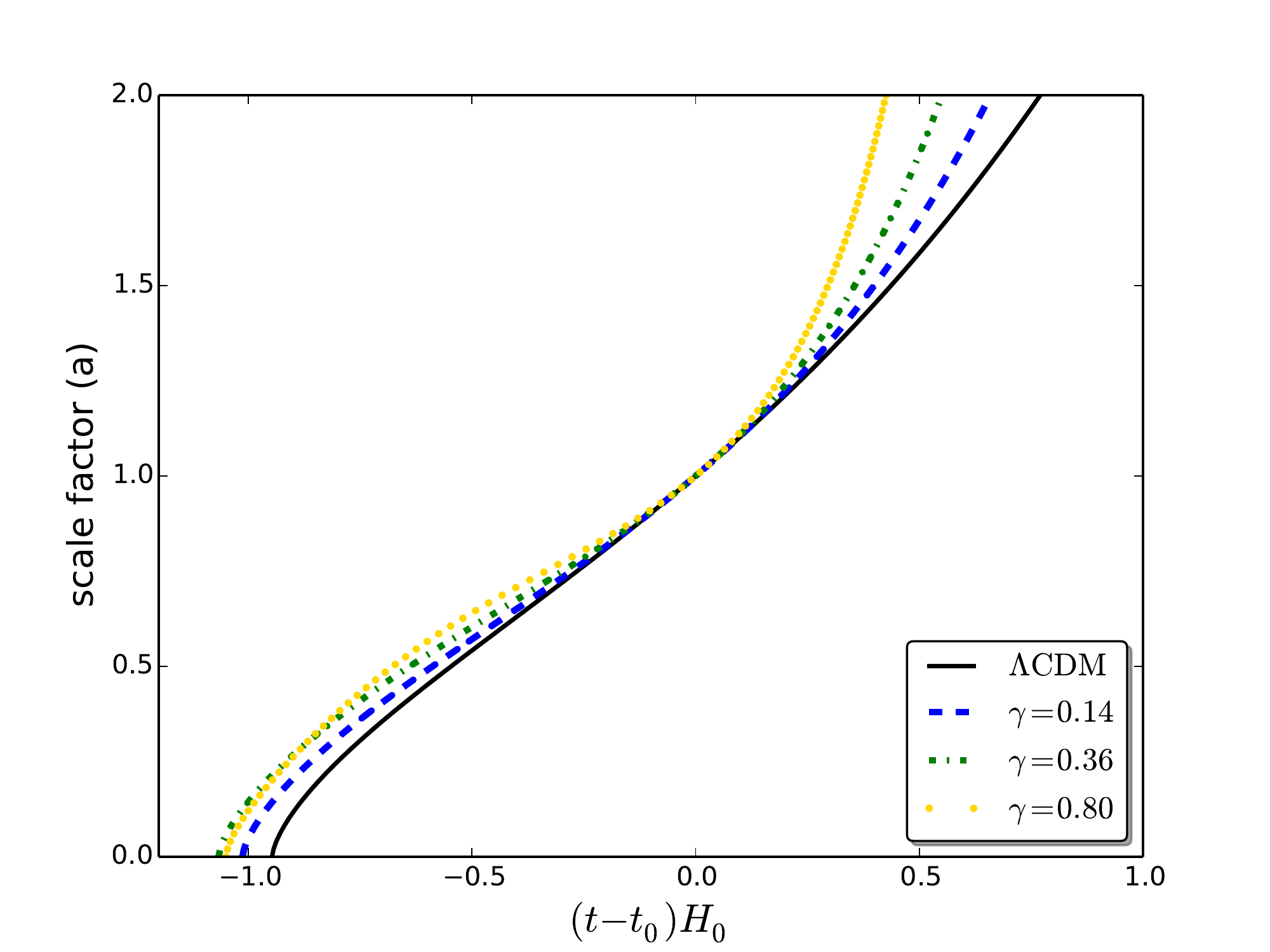}
  \caption{Scale factor as a function of $t-t_0$ for various values of viscous coefficient.  Other free parameters have been fixed by SNIa observational constraint.  $\Lambda$CDM best fit is given by {\it Planck} observation.}
  \label{fig:at}
\end{figure}
In Fig. \ref{fig:at} we computed scale factor as a function of $t-t_0$ for various values of viscosity for viscous dark energy model. Increasing the value of $\gamma$ leads to increase the age of Universe, but when $\gamma$ is grater than a typical value ($\gamma_{\times}$), the dominant behavior of viscose dark energy is similar to a Quintessence model (see Fig. \ref{fig:at}).

 In the next section, we propose an \emph{Action} in the scalar-tensor theories which its equation of motions has the same behavior in our viscous dark energy model.

\section{Corresponding Action of model}


As mentioned in introduction, to explain the late time accelerating expansion of the Universe, dynamical dark energy including field theoretical orientation and phenomenological dark fluids \cite{Copeland:2006wr,Amendola:2012ys}, modified general relativity  \cite{Horndeski:1974wa,Amendola:2012ys,Copeland:2012qf,Charmousis:2011bf,Appleby:2012rx} and thermodynamics motivated frameworks have been considered in \cite{Kamenshchik:2001cp,Fabris:2001tm,Bento:2002ps}.There is no consensus on where to draw the line between mentioned categories \cite {Amendola:2012ys}. According to scalar field point of view, one can assume that the cosmos has been filled by a phenomenological scalar field generating an accelerated phase of expansion without the need of a specific equation of state for an exotic matter. 
In the previous section, we introduced a bulk-viscous fluid model for dynamical dark energy and calculated some cosmological consequences. There are several proposals to describe the bulk viscosity in the Universe. As an example \textit{superconducting string } can create the viscosity effect for dark fluid \cite{Ostriker_1986}. Particle creation in the Universe may cause to an effective viscosity for vacuum \cite{Hu:1986jd}. Here based on theoretical Lagrangian orientation,
{we will apply a robust method to determine evolution equation of  scalar field, to make a possible correspondence between scalar field and viscosity of dynamical dark energy.}
Following the research done by L. Amendola \cite {Amendola:1993uh},
many models for NMDC have been proposed to investigate inflationary epoch and late time
accelerating expansion for the Universe
\cite{Amendola:1993uh,Germani:2010gm,Granda:2010hb}.
S.~V.~Sushkhov showed that for a specific value of coupling
constants, one can construct new exact cosmological solution without
considering a certain form for scalar field potential
\cite{Sushkov:2009hk,Saridakis:2010mf}. In this section we shall
propose a non-minimal derivative coupling scalar field as a correspondence to  our
dynamical dark energy model. In our viscous dark energy model, it is
possible to have Phantom crossing, therefore one of the proper
actions for describing this dynamical dark energy is an action
containing a Lagrangian with NMDC scalar field
\cite{Perivolaropoulos:2005yv,Gao:2010vr, Banijamali:2012hn}.

{{\subsection{Field equations}}

We consider following action containing a Lagrangian with NMDC scalar field \cite{Sushkov:2009hk, Gumjudpai:2015vio}:

\begin{eqnarray}\label{NMDC_action}
\mathcal{S}=\int{d^{4}x\sqrt{-g}\left(\frac{\rm{M^{2}_{\rm{Pl}}}}{2} R-\frac{1}{2}\left(\epsilon g^{\mu\nu}+\kappa G^{\mu\nu}\right)\partial_{\mu}\phi \partial_{\nu}\phi\right)}+\mathcal{S}_{\rm m},\nonumber\\
\end{eqnarray}

where $R$, $\rm{M_{\rm{Pl}}} $ and $\kappa$  are Ricci scalar, reduced Planck mass and coupling constant between scalar field and Einstein tensor, respectively.  { In the mentioned action $\epsilon$ is $+1$ for Quintessence and $-1$ for Phantom scalar fields and $\mathcal{S}_{\rm m}$ is pressure-less dark matter action.} This class of actions with different values for couplings to the curvature corresponds to low energy limit of some higher dimensional theories such as superstring \cite{Appelquist_1983,Holden:1999hm,Easson:2006jd} and quantum gravity \cite{Uzan:1999ch}. We suppose zero potential for scalar field to get rid of any fine-tuned potentials \cite{Sushkov:2009hk}. Varying the action in Eq.~(\ref{NMDC_action}) with respect to the metric tensor and the scalar field, leads to field equations \cite{Sushkov:2009hk,Nozari:2016ilx}. Energy-momentum tensor of NMCD field, $T^{\phi}_{\mu \nu}$ is obtained by variation of action (\ref{NMDC_action}) respect to the metric tensor, $g_{\mu \nu}$ and it is:
\begin{eqnarray}
T^{\phi}_{\mu \nu}=\epsilon \Theta_{\mu \nu }+\kappa \Pi_{\mu \nu},
\end{eqnarray}
where we defined $\Theta_{\mu \nu}$ and $\Pi_{\mu \nu}$ according to:
\begin{eqnarray}
\Theta_{\mu \nu}&\equiv&-\dfrac{1}{2} g_{\mu \nu}(\nabla \phi)^2+ \nabla_{\mu}\phi\nabla_{\nu}\phi,
\end{eqnarray}
and
\begin{eqnarray}
\Pi_{\mu \nu}&\equiv&-\frac{1}{2}R_{\mu \nu}\left(\nabla\phi  \right)^2+\frac{1}{2}g_{\mu \nu}(\Box \phi)^2\nonumber\\
&-&\frac{1}{2}g_{\mu \nu}G_{\alpha \beta}\nabla^{\alpha}\phi\nabla^{\beta}\phi-\frac{1}{2}g_{\mu \nu}R_{\alpha \beta}\nabla^{\alpha}\phi\nabla^{\beta}\phi \nonumber \\
& + & R_{\mu \alpha \nu \beta}\nabla^{\alpha}\phi\nabla^{\beta}\phi-\frac{1}{2}g_{\mu\nu}\nabla_{\beta}\nabla_{\alpha} \phi \nabla^{\beta}\nabla^{\alpha}\phi\nonumber\\
&+&G_{\nu \alpha}\nabla^{\alpha}\phi\nabla_{\mu}\phi + G_{\mu \alpha}\nabla^{\alpha}\phi\nabla_{\nu}\phi +\frac{1}{2}R\nabla_{\mu}\phi\nabla_{\nu}\phi \nonumber\\
&+&\nabla_{\mu}\nabla^{\alpha}\phi\nabla_{\nu}\nabla_{\alpha}\phi-\Box\phi\nabla_{\nu}\nabla_{\mu}\phi.
\end{eqnarray}
{As mentioned before, by taking the variation of action represented by Eq.(\ref{NMDC_action}), general expression for field equation is retrieved.  Now we can determine the $00$ and $11$ components of this equation, we find the first and second Friedmann equations for mentioned NMDC action with standard model for dark matter read as:
\begin{eqnarray}\label{eq:first_order_diff}
H^2=\frac{8\pi G_N}{3} \left[\rho_m+\frac{\dot{\phi}^2}{2}\left (\epsilon-9\kappa H^2   \right) \right]
\end{eqnarray}
\begin{eqnarray}\label{eq:second_friedmann_eq}
-2\dot{H}-3H^2 =4\pi G_N \left [\epsilon+\kappa \left (2\dot{H}+3 H^2+4H\frac{\ddot{\phi}}{\dot{\phi}}         \right )       \right ]
\end{eqnarray}
To find differential equation for scalar field, we should take variation in Eq. (\ref{NMDC_action}) with respect to the scalar filed considering FRW background metric, namely:
\begin{eqnarray}\label{eq:second_order_diff}
\epsilon(\ddot{\phi}+3H\dot{\phi})-3\kappa\left [ H^2\ddot{\phi}+2H\dot{H}\dot{\phi}+3H^3\dot{\phi}      \right ]=0.
\end{eqnarray}
These equations aren't independent. By combining the Eq.~(\ref{eq:second_order_diff}) and Eq.~(\ref{eq:second_friedmann_eq}) one can get Eq.~(\ref{eq:first_order_diff}).}

\subsection{Bulk viscose solution}
{
 Hereafter, we are looking for finding consistent solutions for coupled differential equations (Eqs. (\ref{eq:first_order_diff}) and (\ref{eq:second_order_diff})) for a given Hubble parameter.  Since, we have no scalar potential,  it is not necessary to impose additional constraint. 
The Hubble parameter for flat bulk viscose dark energy model is:
\begin{eqnarray}\label{eq:hubble23}
H^2=H_0^2\left[\Omega_{\rm m}^0 a^{-3}+\Omega_{\rm DE}^0\left(1 +\frac{9 \gamma}{2\sqrt{\Omega^{0}_{\rm DE}}}\ln{a}\right)^{2}\right],
\end{eqnarray}
Therefore, we try to find special solution of Eq.~(\ref{eq:first_order_diff}) which its Hubble parameter 
 is represented by Eq. (\ref{eq:hubble23}). 
It turns out that coupling parameter between Einstein tensor and the kinetic term is related to the viscosity of dynamical dark energy. In the absence of viscosity coefficient, coupling parameter would be vanished and standard minimal action to be retrieved. 
 According to mentioned explanation,
Eq. (\ref{eq:first_order_diff}) takes the following form:
\begin{widetext}
\begin{eqnarray}
\left( \frac{d \tilde{\phi}}{da} \right) ^{2}=-
 \frac{\Omega^{0}_{\rm{DE}}\left( 1- \frac{\ln{a}}{\ln{\tilde{a}}}                     \right)^{2} }{a^{2}\left( \Omega_{\rm{m}}^{0}a^{-3}+\Omega^{0}_{\rm{DE}}\left( 1- \frac{\ln{a}}{\ln{\tilde{a}}}                     \right)^{2}\right)              \left\lbrace  \epsilon - 9\kappa H_0^2
 \left[\Omega_{\rm{m}}^{0}a^{-3}+\Omega^{0}_{\rm{DE}}\left( 1- \frac{\ln{a}}{\ln{\tilde{a}}}    \right)^{2}                             \right]\right\rbrace},
\end{eqnarray}
\end{widetext}
where we define $ \tilde{\phi}\equiv\sqrt{\dfrac{4\pi G_{N} }{3}} \phi$. To avoid possible singularity in above equation and construct a {\it ghost free Action}, one can find following relation between $\epsilon$ and $\kappa$:}
\begin{eqnarray}\label{kappap}
  \kappa =\frac{\epsilon}{9H_{0}^{2}(1-\Omega^{0}_{\rm{DE}})}\exp\left(- \frac{2 \sqrt{\Omega^{0}_{\rm{DE}} }}{3\gamma}           \right).
\end{eqnarray}
In the upper panel of Fig. \ref{fig:field_3}, we plot $\dot{\phi}^2$ as a function of scale factor for $\epsilon=\pm1$. Due to the functional form of dynamical dark energy model and to ensure that  scalar field to be a real quantity, therefore, we should take $\epsilon=-1$ causing the coupling coefficient becomes negative. Lower panel of Fig. \ref{fig:field_3} indicates $\tilde{\phi}$ versus scale factor for best fit values of parameters based on JLA catalog (see next sections). At the early epoch the contribution of this scalar field as a model of dynamical dark energy is ignorable while on the contrary at the late time, it affects on the background evolution considerably.
\begin{figure}
\centering
  \includegraphics[width=.9\columnwidth]{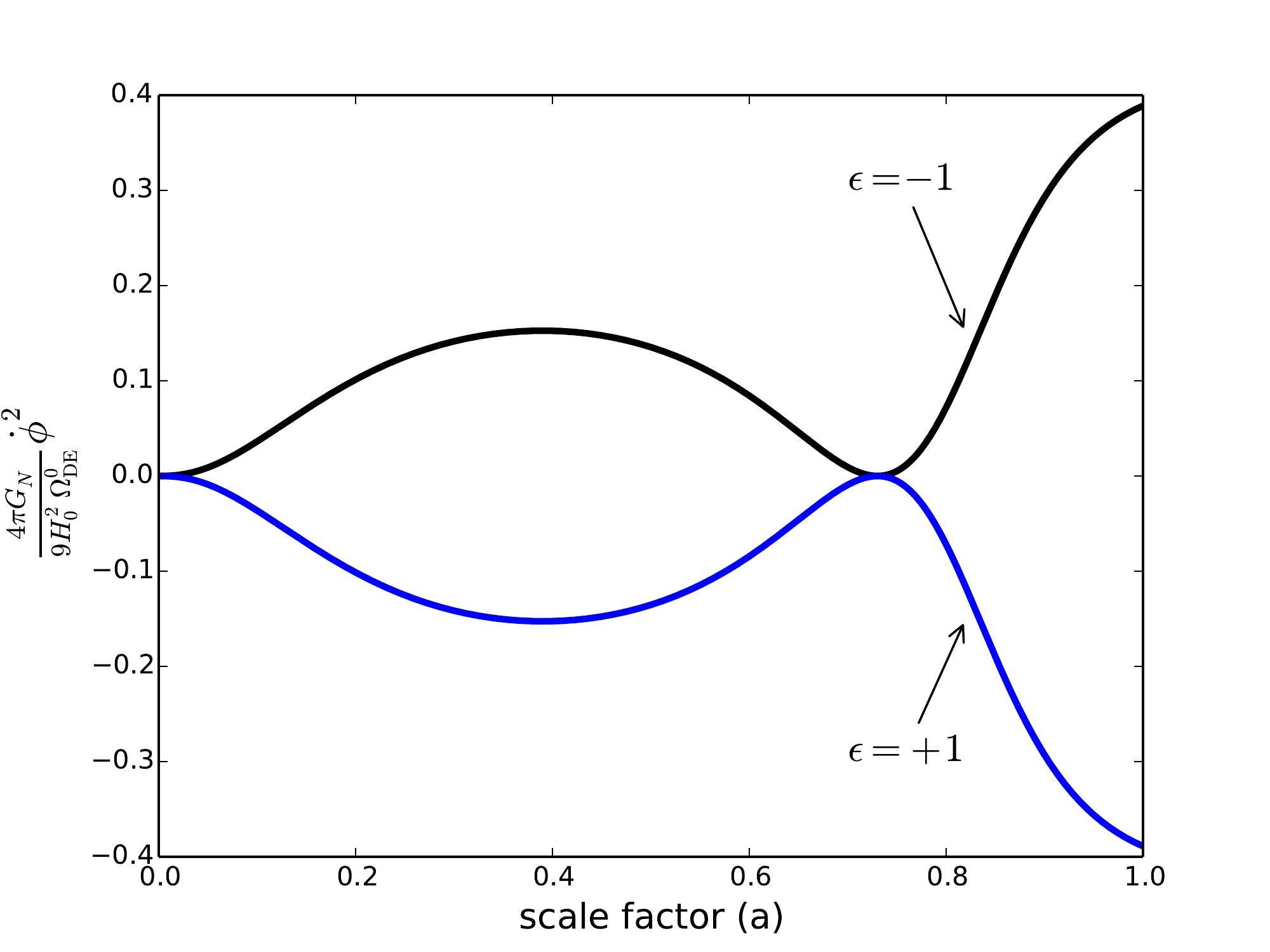}\\
   \includegraphics[width=.9\columnwidth]{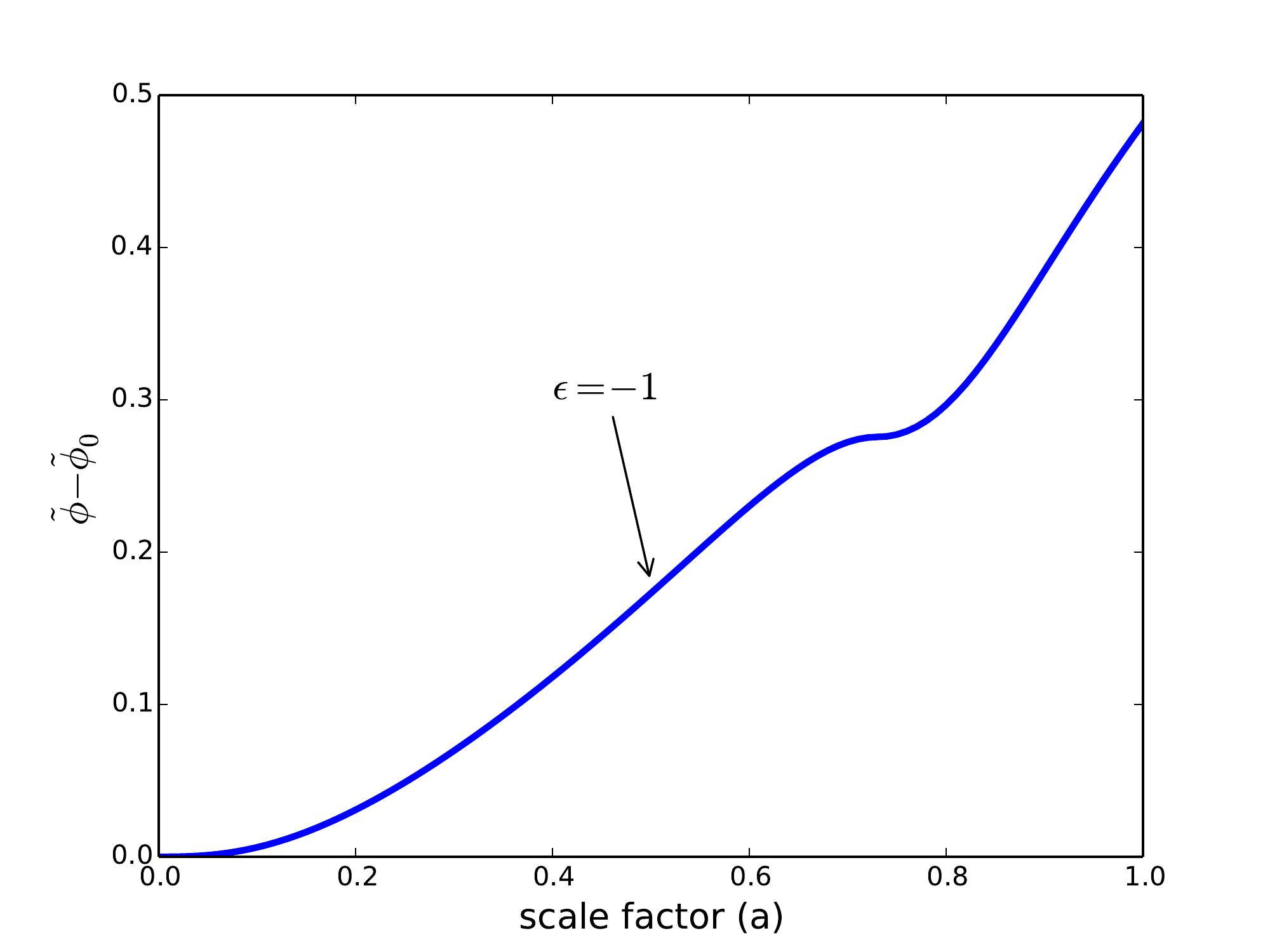}
  \caption{\textit{Upper panel:} Evolution of field $\dot{\phi}^{2}$ as a function of scale factor for different value of parameter $\epsilon$.  \textit{Lower panel:} Evolution of field $\tilde{\phi}$ as a function of scale factor. Here $\tilde{\phi}_0$ is the initial condition for scalar field. The other free parameters have been fixed by SNIa observational constraint}
  \label{fig:field_3}
\end{figure}


In the next section we will use most recent and precise observational data sets to put constraints on free parameters of our model and to evaluate the consistency of our dynamical dark energy model. Also we will rely on a reliable geometrical diagnostic which is so-called $Om$ measure to do possible discrimination between our model and $\Lambda$CDM. New observable quantity which is so-called  cosmographic distance ratio which is free of bias effect is also considered for this purpose \cite{Jain:2003tba,Taylor:2006aw,Kitching:2008vx}.

\section{Effect on geometrical parameters }
In this section, the effect of viscous dark energy model on the geometrical parameters of Universe will be examined. We consider comoving distance, apparent angular size (Alcock-Paczynski test), comoving volume element, the age of Universe, cosmographic parameters and $Om$ diagnostic.

According to theoretical setup mentioned in section II, the list of free parameters of underlying is as follows:
$$ \{ {\Theta _p}\} :\gamma , \Omega _{\rm DE}^0,{\Omega _b}{h^2},{\Omega _{\rm m}}{h^2},{H_0},\tau_{\it opt} ,{\mathcal{A}_s},{n_s} $$
where $A_s$ is the scalar power spectrum  amplitude. Also  $n_s$ is corresponding to exponent of mentioned power spectrum. $\Omega _b{h^2}$ and ${\Omega _{\rm m}}{h^2}$ are dimensionless baryonic and cold dark  matter energy density, respectively. The optical depth is indicated by $\tau_{\it opt}$.
\subsection{Comoving Distance}
The radial Comoving distance for an object located at redshift $z$ in the FRW metric reads as:
\begin{equation}
r(z;\{ {\Theta _p}\})  =  \int_0^z {dz'\over
H(z';\{ {\Theta _p}\} )}, \label{comoving},
\end{equation}
here $H(z';\{ {\Theta _p}\} )$ is given by Eq. (\ref{eq:hubble}). As
indicated in Fig. \ref{comoving}, by increasing the value of
$\gamma$ when other parameters are fixed, the contribution of
viscous dark energy becomes lower than cosmological constant.
Therefore,  the comoving distance is longer than that of in
$\Lambda$CDM or Quintessence models. For
$\gamma>\gamma_{\times}\simeq 0.36$, our result demonstrates  a
crossover in behavior of comoving distance due to changing the role
of viscous dark energy from Phantom to Quintessence class (Fig. \ref{comoving}).

\begin{figure}
\centering
  \includegraphics[width=.9\columnwidth]{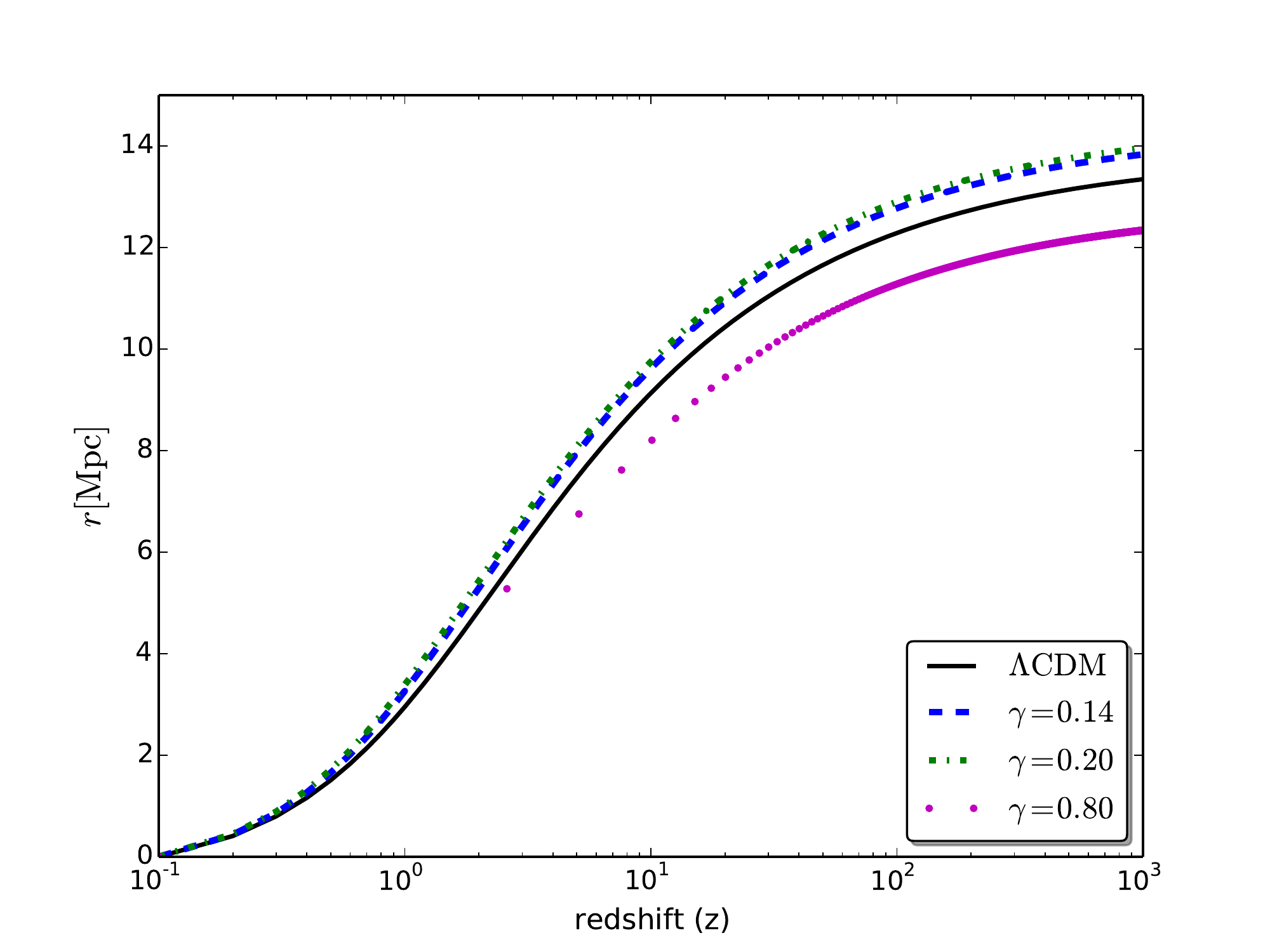}
    \caption{The effect of viscosity on the radial comoving distance in FRW metric. The other free parameters have been fixed by SNIa observational constraint.  $\Lambda$CDM best fit is given by {\it Planck} observation. }
  \label{comoving}
\end{figure}
\subsection{Alcock-Paczynski test}
The so-called Alcock-Paczynski test is another interesting probe for dynamics of background based on anisotropic clustering and it doesn't depend on the evolution of the galaxies . By measuring
the angular size in different redshifts in isotropic rate of expansion case, one can write \cite{Alcock:1979mp}:
\begin{equation}
{\Delta z\over \Delta \theta} =
H(z;\{ {\Theta _p}\})r(z;\{ {\Theta _p}\}) \label{alpa}.
\end{equation}
We should point out that one of advantages of Alcock-Paczynski test is that
it is independent of standard candles as well as evolution of galaxies \cite{Lopez-Corredoira:2013lca}. 
 Considering the evolution of cosmological objects radius along the line of sight  to the same in perpendicular to line of sight ratio, a complete representation of mentioned quantity reads as:
 \begin{equation}
\mathcal{Y}(z;\{\Theta_p\})=\frac{z+1}{z} H(z;\{ {\Theta _p}\})d_A(z;\{ {\Theta _p}\}), \label{alpa}
\end{equation}
here $d_A$ is angular diameter distance. The observed value for $\mathcal{Y}$ at three redshifts are $\mathcal{Y}(z=0.38)=1.079\pm0.042$, $\mathcal{Y}(z=0.61)=1.248\pm0.044$ \cite{Alam:2016hwk} and $\mathcal{Y}(z=2.34)=1.706\pm0.083$ \cite{Melia:2015cqa}. The upper panel of Fig.~\ref{alc} represents $\Delta z/ \Delta \theta$ as a function of redshift.  We normalized this value to the $\Lambda$CDM model ($H(z;\gamma=0)r(z;\gamma=0) $) constraining by {\it Planck} observations. By increasing the contribution of viscosity when other parameters are fixed, we find that our model has up and down behavior at low redshift  where almost transition from dark matter to dark energy era occurred. The lower panel of  Fig.~\ref{alc} indicates observable values of Alcock-Paczynski for various viscosity coefficients. As illustrated by this figure, increasing the value of viscosity leads to better agreement with observed data for low redshift while there is a considerable deviation for high redshift.
\begin{figure}
\centering
  \includegraphics[width=.9\columnwidth]{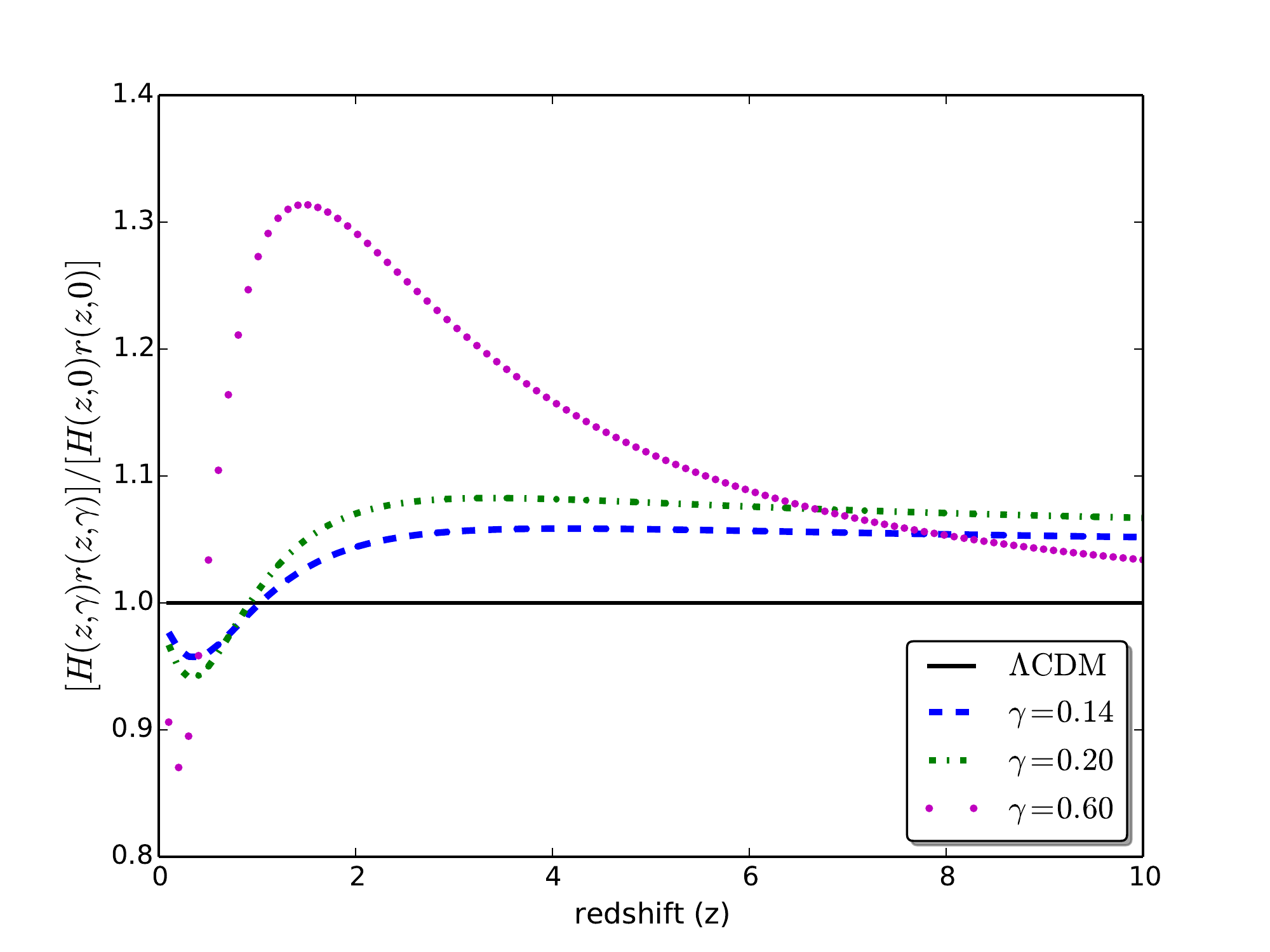}\\
  \includegraphics[width=.9\columnwidth]{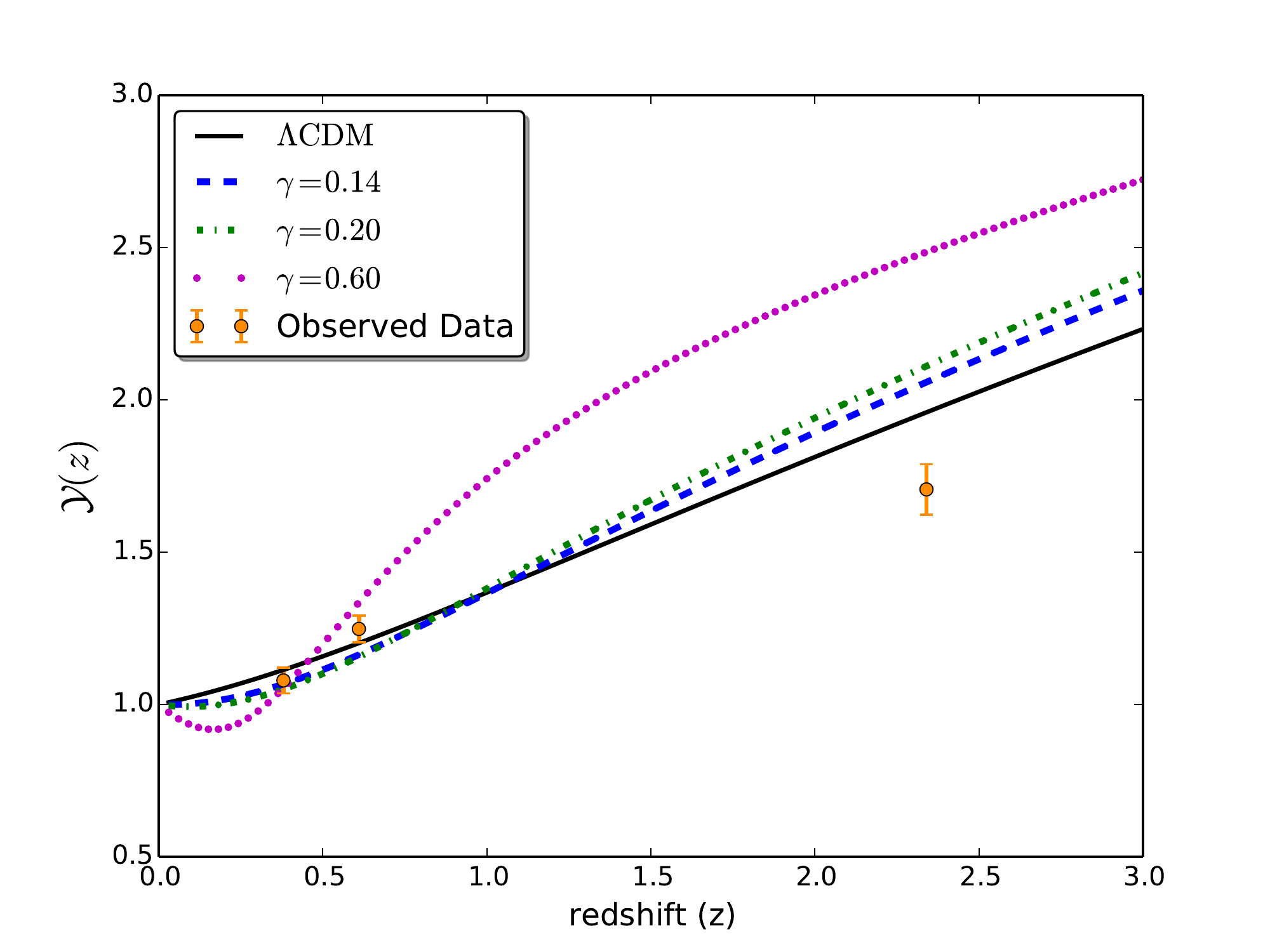}
    \caption{{\it Upper panel}: Alcock-Paczynski test, compares $\Delta z/{\Delta \theta}$
normalized to the case of $\Lambda$CDM model $(\gamma=0)$as a function of
redshift for different viscosity coefficients. {\it Lower panel}:  $\mathcal{Y}(z;\{\Theta_p\})$ for viscous dark energy and $\Lambda$CDM models. In addition three observed points are illustrated for better comparison. The other free parameters have been fixed by SNIa observational constraint.  $\Lambda$CDM best fit is given by {\it Planck} observation.}
  \label{alc}
\end{figure}


\subsection{Comoving volume element}
An other geometrical parameter is the comoving volume element used in number-count tests such as lensed quasars, galaxies, or clusters of galaxies. Mentioned quantity is written in terms of comoving distance and Hubble parameters as follows:
\begin{equation}
f(z;\{ {\Theta _p}\}) \equiv {dV\over dz d\Omega} =\frac{r^2(z;\{ {\Theta _p}\})}{H(z;\{ {\Theta _p}\})}.
\end{equation}
Refereing to Fig. \ref{fig:cve}, one can conclude that the comoving
volume element becomes maximum around $z\simeq 2.6$ for $\Lambda$CDM. In the bulk viscous
model for $\gamma=0.14$ the maximum occurs  at redshift around $z\simeq 2.3$. For
larger value of $\gamma$ exponent, the position of this maximum
shifts to the lower redshifts corresponding to the case with lower
contribution of viscous dark energy, as indicated in
Fig.~\ref{fig:density-ratio-a}.

\begin{figure}
\includegraphics[width=.9\columnwidth]{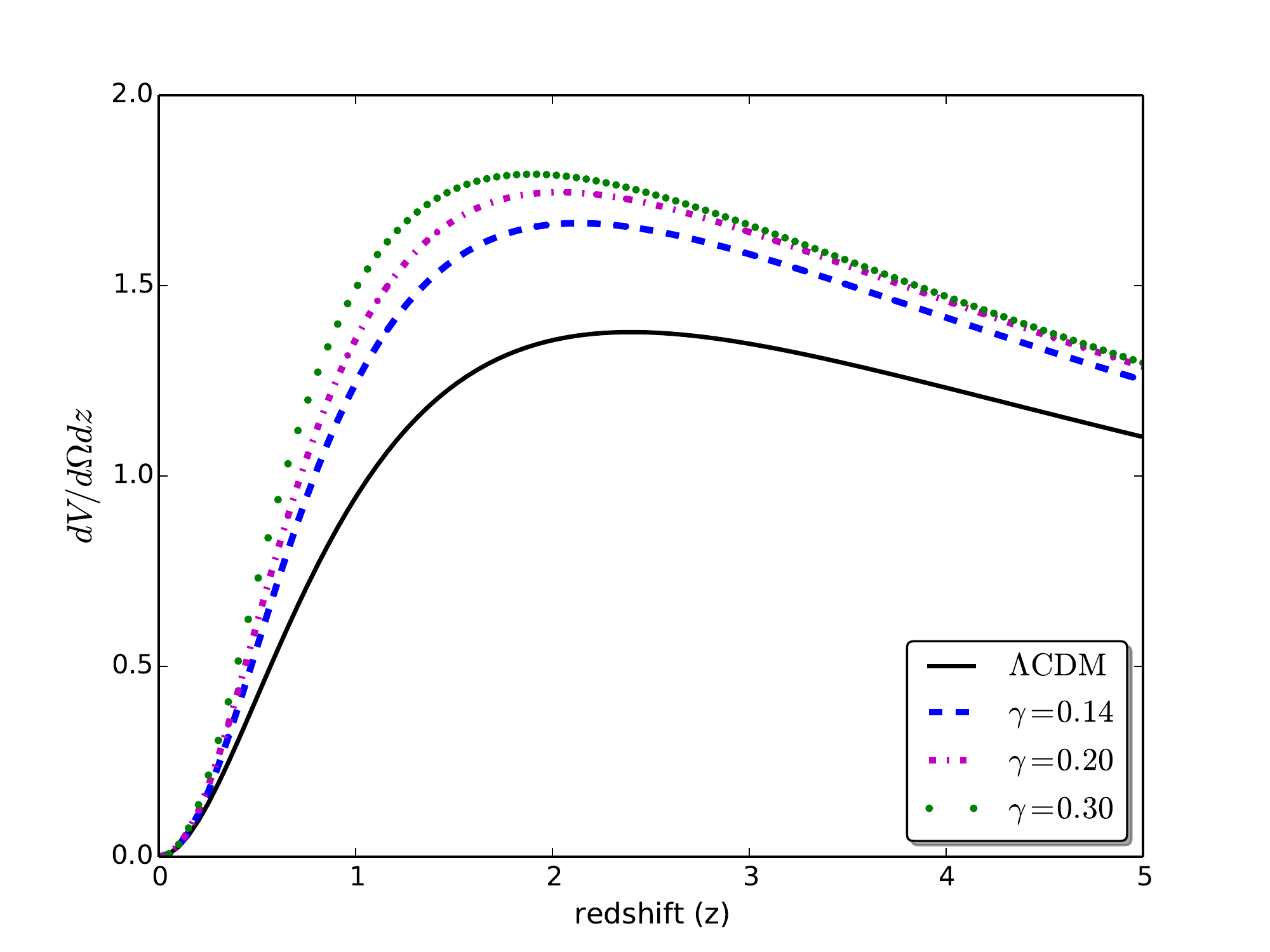}
\centering
 \caption{The
comoving volume element versus redshift for various values of $\gamma$
exponent. Increasing $\gamma$ shifts the position of maximum value
of the volume element to the lower redshifts. The other free parameters have been fixed by SNIa observational constraint.  $\Lambda$CDM best fit is given by {\it Planck} observation. } \label{fig:cve}
 \end{figure}

\subsection{Age of Universe}
An other interesting quantity is the age of Universe computed in a cosmological model.
The age of Universe at given redshift can be computed by integrating from the big bang indicated by infinite redshift up to $z$ as:
\begin{equation}\label{age}
t(z; \{\Theta_p\}) =  \int_z^\infty {dz'\over
(1+z')H(z';\{\Theta_p\})},
\end{equation}
To compare age of Universe we set the lower value of integration to zero and we represent this quantity by $t_0$. We plotted $H_0t_0$ (Hubble parameters times the age of Universe) as a function of $\gamma$ for values of cosmological parameters constrained by JLA observation in Fig.~\ref{fig:age}. The age of Universe has a maximum value for $\gamma_{\times}\simeq0.36$. 
This behavior is due to the dynamical nature of our viscous dark energy model. Namely, for lower values of $\gamma<\gamma_{\times}$, viscous dark energy model is almost categorized in Phantom class during the wide range of scale factor while underlying dark energy model is devoted to Quintessence class by increasing the value of $\gamma>\gamma_{\times}$.
The existence of dark energy component is also advocated by "age crises" (for full review of the cosmic age see \cite{Julian:1967zz}). In the next section, we will check the consistency of our viscous dark energy model according to comparing the age of Universe computed in this model and with the age of old high redshift galaxies located in various redshifts.

\begin{figure}
\includegraphics[width=.9\columnwidth]{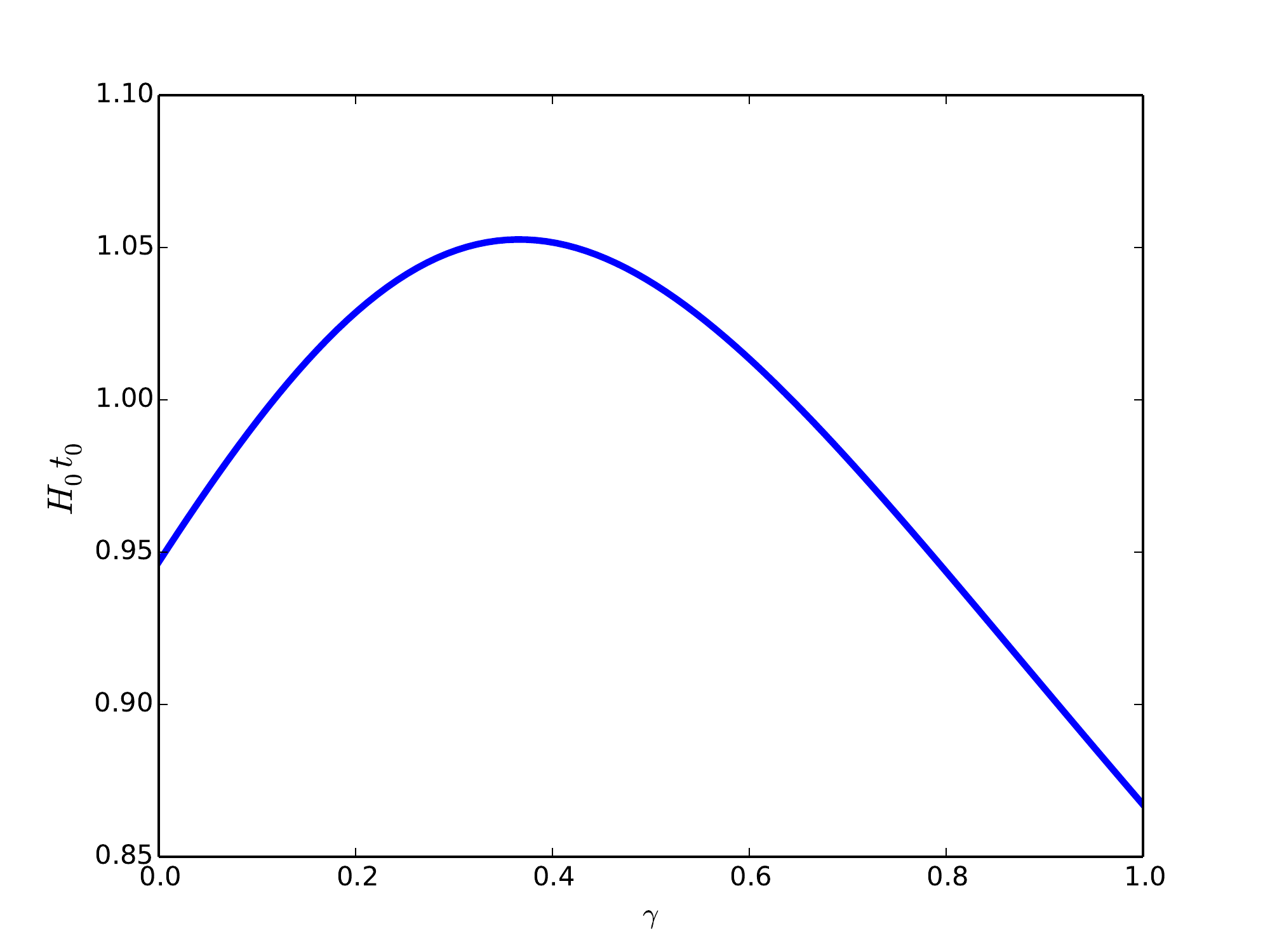}
\centering
\caption{
The quantity $H_0t_0$ (age of Universe times the Hubble constant at the present
time) as a function of $\gamma$.  In our dark energy model, age of Universe has a maximum at $\gamma_{\times}\simeq0.36$ with maximum value equates to $(H_0t_0)_{max}=1.052$. We chose other cosmological parameters in flat Universe with observational constraint using JLA catalog.}
\label{fig:age}
 \end{figure}

 \subsection{Cosmographic parameters}
One of the most intriguing questions concerning the late time accelerating expansion in the Universe is that, the possibility of distinguishing  between cosmological constant and dark energy models. There are many attempts in order to answer mentioned question \cite{Sahni:2008xx,Shafieloo:2014ypa}. High sensitivity as well as model independent properties should be considered for introducing a reliable diagnostic measure. Using cosmography parameters, we are able to study some kinematic properties of viscous dark energy model.
First and second cosmography parameters are Hubble and deceleration parameters. Other kinematical parameters defined by \cite{Dunsby:2015ers}:
\begin{eqnarray}
j \equiv \frac{1}{a\, H^{3}} \frac{d^{3}a}{dt^{3}},\,\,\,\,
s \equiv \frac{1}{a\, H^{4}} \frac{d^{4}a}{dt^{4}},\\ \nonumber
l \equiv  \frac{1}{a\, H^{5}} \frac{d^{5}a}{dt^{5}},\,\,\,\,
m \equiv \frac{1}{a\, H^{6}} \frac{d^{6}a}{dt^{6}}.
\end{eqnarray}
these parameters are called \textit{jerk}, \textit{snap}, \textit{lerk} and \textit{maxout}, respectively. It is worth noting that  these parameters are not independent together and they are related to each other by simple equations. If we denote the derivative respects to the cosmic time by \textit{dot}, we can write:
\begin{eqnarray}
\dot{H}&=&-H^{2}\left(1+q\right),\nonumber\\
\ddot{H}&=&H^{3}\left( j+3 q+2 \right),\nonumber\\
\dddot{H}&=&H^{4}\left[ s-4 j-3 q \left( q+4\right)-6\right], \nonumber \\
\ddddot{H}&=&H^{5}\left[l-5 s+10 \left(q+2\right)j+30\left(q+2\right)q+24  \right], \nonumber \\
H^{(5)}&=&H^{6}\{ m-10j^{2}-120 j \left( q+1\right)\nonumber \\
&&-3\left[2l+5\left(24 q+18q^{2}+2q^{3}-2s-qs+8\right)\right]\}. \nonumber\\
\end{eqnarray}
In Figs. \ref{fig:qz} we show $\dot{a}$, $q(z)$, \textit{lerk}, \textit{snap}, \textit{lerk} and \textit{maxout} cosmographic parameters of our model in comparison to $\Lambda \rm{CDM}$ model. As indicated in mentioned figures at low redshift there is a meaningful difference between bulk-viscous model and $\Lambda \rm{CDM}$ model.

\begin{figure}
\centering
\includegraphics[width=.9\columnwidth]{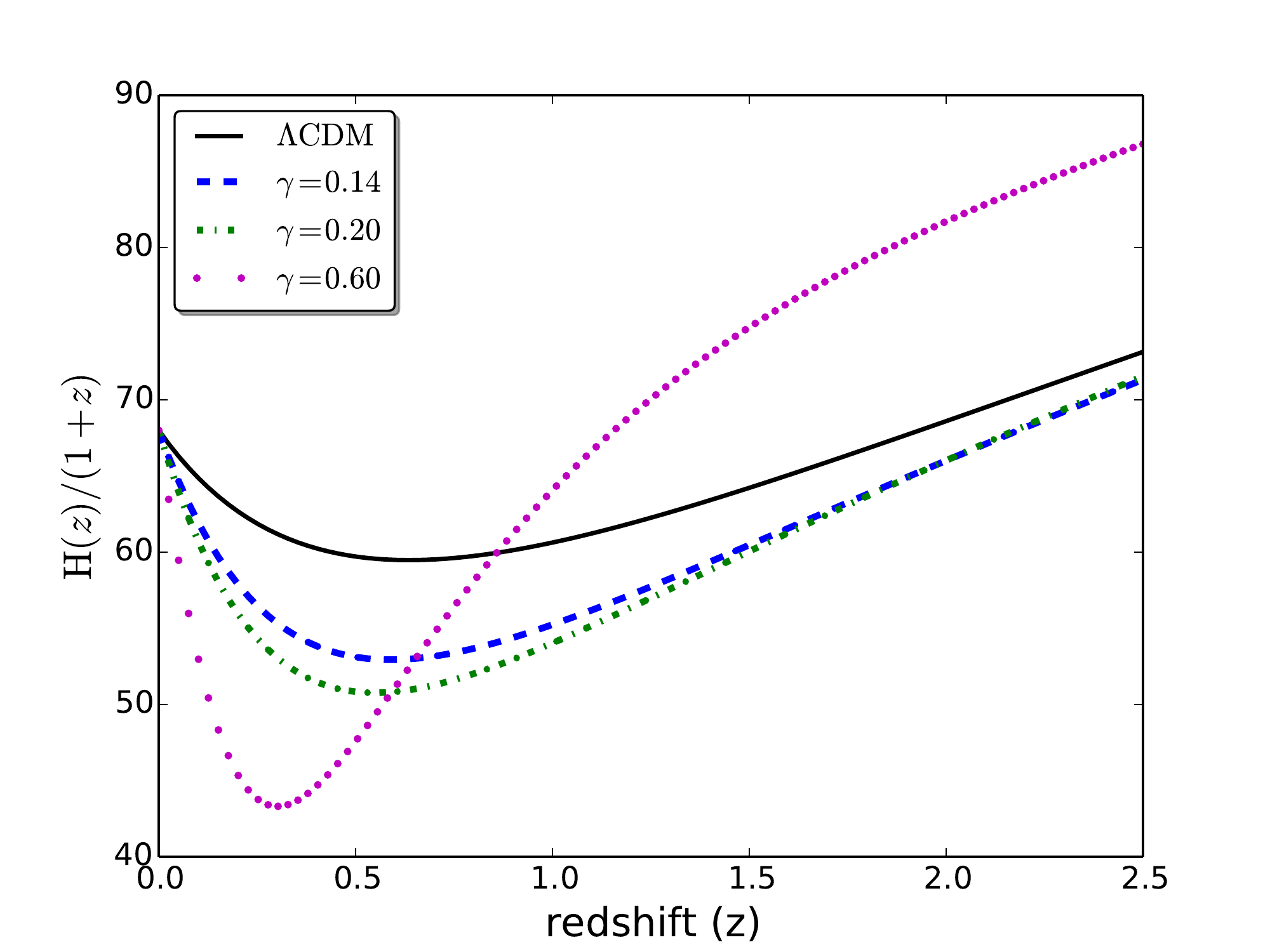}\\
  \includegraphics[width=.9\columnwidth]{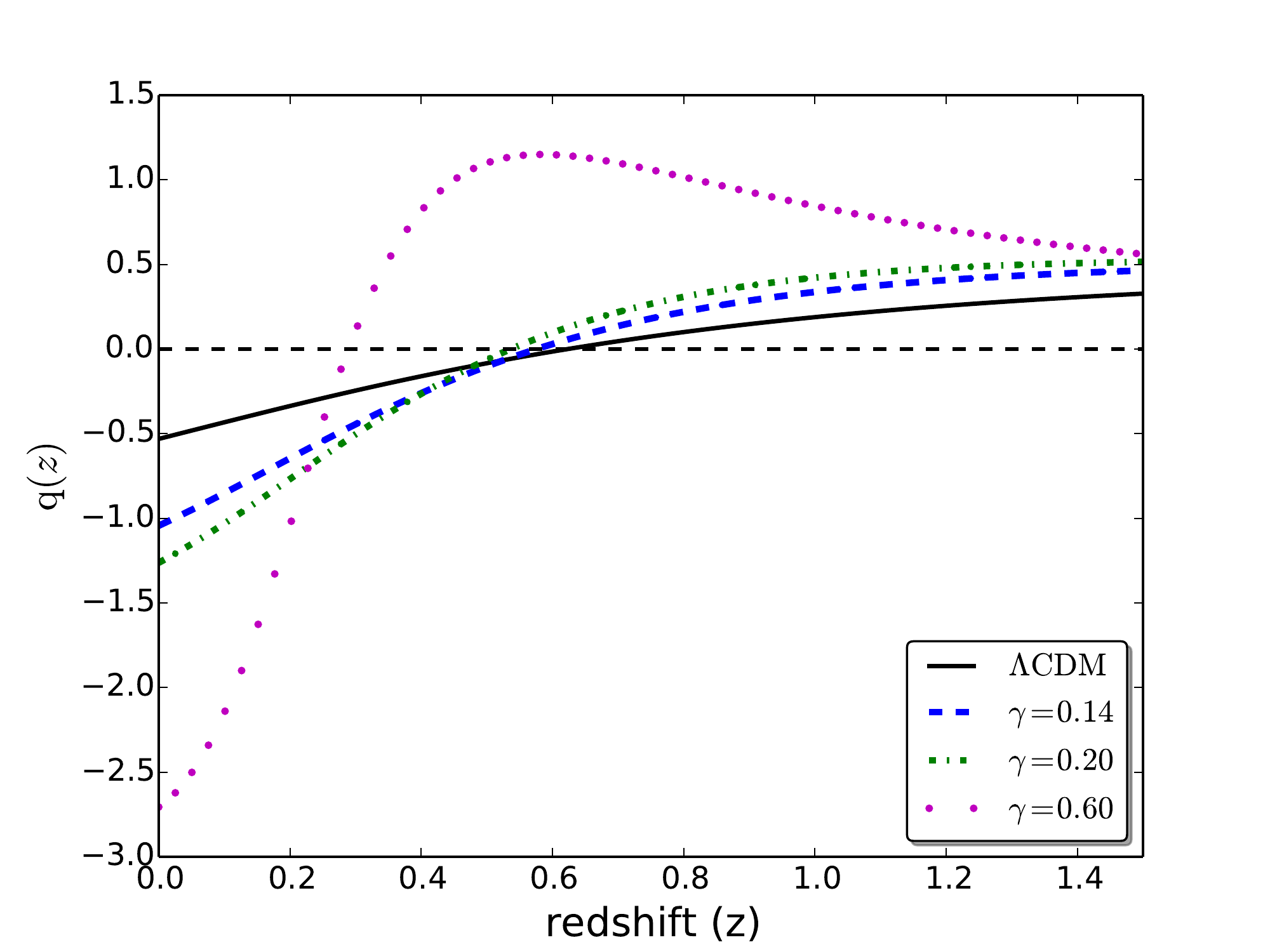}\\
  \includegraphics[width=1.1\columnwidth]{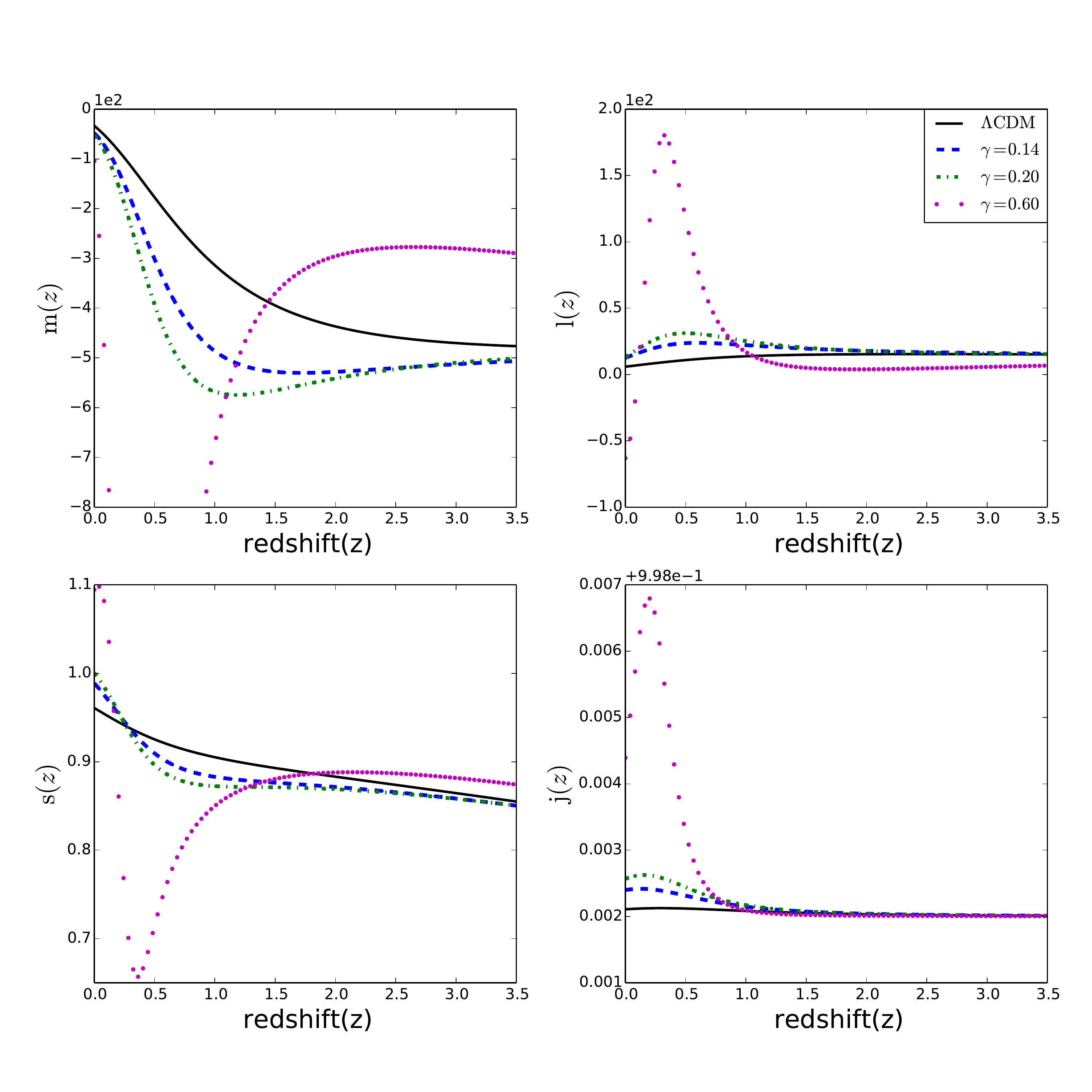}
  \caption{ {\it Upper panel:} The  $\frac{H(z)}{(1+z)}$ as a function of redshift. {\it Middle panel:} Deceleration probe diagnostic. {\it Lower panel:} {\it jerk}, {\it snap}, {\it lerk} and {\it maxout} parameters for bulk viscous model with respect to $\LCDM$ model. Solid lines represents corresponding quantity for $\LCDM$. Other lines are associated with different values for viscosity. The rest of free parameters have been fixed according to JLA observation at 1$\sigma$ confidence level.}
  \label{fig:qz}
\end{figure}


\subsection{$\mathit{Om}$ diagnostic}
The $Om$ diagnostic method is indeed a geometrical diagnostic which combines Hubble parameter and redshift. It can differentiate dark energy model from $\Lambda \rm{CDM}$. An other measure is acceleration probe. Sahni and his collaborators demonstrated that, irrespective to matter density content  of Universe, acceleration probe can discriminate various dark energy models  \cite{Sahni:2008xx}. $\mathit{Om}(z)$ diagnostic for our spatially flat Universe reads as:
\begin{equation}
Om(z;\{\Theta_P\}) \equiv \frac{{\mathcal{H}}^2(z;\{\Theta_P\}) -1}{(1+z)^3-1}.
\end{equation}
where ${\mathcal{H}}\equiv\frac{H}{H_0}$ and $H$ is given by Eq. (\ref{eq:hubble}).
For $\LCDM$ model $Om(z)=\Omega_m$ while for other dark energy models,  $Om(z)$ depends on redshift \cite{Shafieloo:2014ypa}. Phantom like dark energy corresponds to the positive slope of $\mathit{Om}(z)$ whereas the negative slope means that dark energy behaves like Quintessence \cite{Shahalam:2015lra}. In addition, $Om(z)$ depends upon no higher derivative of the luminosity distance in comparison for $w(z)$ and the deceleration parameter $q(z)$, therefore, it is less sensitive to observational errors \cite{Sahni:2008xx}.  Another feature of $\mathit{Om}(z)$ is that the growth of $\mathit{Om}(z)$ at late time favours the decaying   dark energy models \cite{Shafieloo:2009ti}.

 Fig. \ref{fig:omz} indicates acceleration probe measure for cosmological constant with different values of equation of states and viscous dark energy model. viscous dark energy model with $\gamma<\gamma_{\times}$ belongs to Phantom like dark energy. In mentioned figure,  solid lines represents $Om(z)$ for cosmological constant. Dashed and dot-dashed lines represents $Om(z)$ for $\Lambda$CDM with $w=-0.90$ and $w=-1.20$, respectively.    Think solid line with corresponding 1$\sigma$ confidence interval determined by JLA observation represents $Om(z)$ for viscous dark energy model for $\gamma=0.14$. Long dashed line corresponds to $\gamma=0.60$ demonstrating that dynamical dark energy model has almost Quintessence behavior during the evolution of Universe.

\begin{figure}
\centering
  \includegraphics[width=.9\columnwidth]{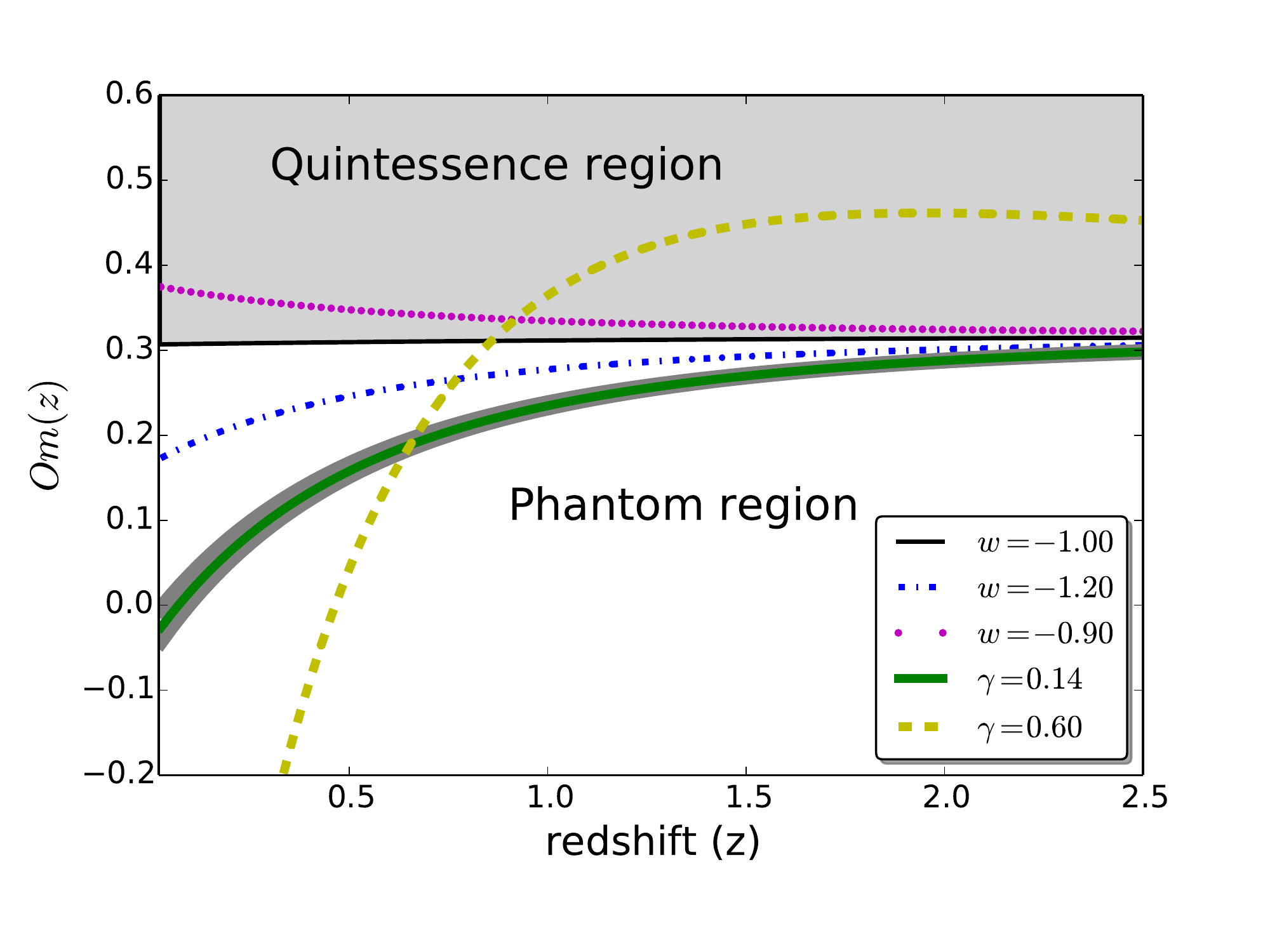}
  \caption{$Om(z)$ diagnostic: Solid lines represents $Om(z)$ for cosmological constant. Dashed and dot-dashed lines represents $Om(z)$ for $\Lambda$CDM with $w=-0.90$ and $w=-1.20$, respectively.    Think solid line with corresponding 1$\sigma$ confidence interval determined by JLA observation represents $Om(z)$ for viscous dark energy model for $\gamma=0.14$. Long dashed line corresponds to $\gamma=0.60$ demonstrating that dynamical dark energy model has almost Quintessence behavior during the evolution of Universe. }
  \label{fig:omz}
\end{figure}

 \section{Consistency with recent Observations} \label{s:constrains}

In this section we consider most recent observational data sets to constrain free parameters of our model. Accordingly, we are able to check the consistency of viscous dark energy model. In principle one can refer to the following observables to examine the nature of dark energy:  \\
I) Expansion rate of the Universe.\\
II) Variation of gravitational potential, producing ISW effect. \\
III) Cross-correlation of CMB and large scale structures.\\
IV) Growth of structures.\\
V) Weak lensing.\\
Throughout this paper we concentrate on almost evolution expansion such as distance modulus of Supernova Type Ia and the Gamma Ray Bursts (GRBs), Baryon acoustic oscillations (BAO) and Hubble Space Telescope (HST). The CMB power spectrum which is mainly affected by background evolution is also considered. The new cosmographic data sets namely cosmographic distance ratio and age test will be used for further consistency considerations. We assume a flat background, so $\Omega^0_{tot} = \Omega^0_{\rm m}+\Omega^0_b+\Omega^0_r+\Omega^0_{\rm DE}=1$. We fixed energy density of radiation by other relevant observations \cite{Ade:2015xua}. The priors considered for parameter space  have been reported in Table \ref{prior}.

\begin{table}[]
\centering
\begin{tabular}{llccc}
\hline
\hline Parameter & Prior & Shape of PDF \\
\hline $\Omega_{tot}^0$ &1.000& Fixed&\\
 $\Omega_bh^2$ &$[0.005-0.100]$& Top-Hat&\\
 $\Omega_{\rm m}h^2$ &$[0.001-0.990]$& Top-Hat&\\
 $\gamma$ &$[0.000-0.200]$& Top-Hat&\\
 $H_0$ &$[40.0-100.0]$& Top-Hat&\\
 $\tau_{\it opt}$ &$[0.01-0.80]$& Top-Hat&\\
 $n_s$ &$[0.800-1.200]$& Top-Hat&\\
 $\ln(10^{10} A_s)$ &$[2.000-4.000]$& Top-Hat&\\
\hline
\hline
\end{tabular}
\caption{\label{prior} Priors on parameter space, used in the posterior analysis in this paper. }
\end{table}

\subsection{Luminosity distance implication}
The Supernova Type Ia  (SNIa) is supposed to be a standard candle in cosmology, therefore we are able to use observed SNIa to determine cosmological distance. SNIa is the main evidence for late time accelerating expansion \cite{Riess:1998cb,Perlmutter:1998np}. Direct observations of SNIa do not provide standard ruler but rather gives distance  modulus defined by:
\begin{eqnarray}
\mu (z;\{\Theta_p\}) &\equiv & m - M \nonumber \\ & =&5\log_{10} \left( {\frac{{{d_L}(z;\{\Theta_p\})}}{{Mpc}}} \right) + 25 ,
\end{eqnarray}
where $m$ and $M$ are apparent and absolute magnitude, respectively. For spatially flat Universe, the luminosity distance defined in above equation reads as:
\begin{eqnarray}
d_L(z;\{\Theta_p\}) = \frac{c}{{{H_0}}}(1 + z)\mathop \int_0^z \frac{dz'}{\mathcal{H}(z';\{\Theta_p\})}.
\end{eqnarray}
In order to compare observational data set with that of predicted by our model, we utilize likelihood function with following $\chi^2$:
\begin{eqnarray}
\chi^2_{SNIa}\equiv \Delta\mu^{\dag}\cdot\mathcal{C}^{-1}_{SNIa}\cdot\Delta\mu,
\end{eqnarray}
where $ \Delta\mu\equiv \mu_{obs}(z)-\mu(z;\{\Theta_p\})$ and $\mathcal{C}_{SNIa}$ is covariance matrix of SNIa data sets. $\mu_{obs}(z)$ is observed distance modulus for a SNIa located at redshift $z$ (Relevant data sets and corresponding covariance is available on website \cite{websnia}). Marginalizing over $H_0$ as a nuisance parameter yields \cite{Ade:2015xua}
\begin{eqnarray}
\chi^2_{SNIa}=\mathcal{M}^{\dag}\cdot\mathcal{C}_{SNIa}^{-1}\cdot\mathcal{M}+\mathcal{A}_{SNIa}+\mathcal{B}_{SNIa},
\end{eqnarray}
where $\mathcal{M}\equiv \mu_{obs}(z)-25-5\log_{10}[H_0d_L(z;\{\Theta_p\})/c]$, and
\begin{eqnarray}
&&\mathcal{A}\equiv -\frac{\left[\sum_{i,j}\mathcal{M}(z_i;\{\Theta_p\})\mathcal{C}^{-1}_{SNIa}(z_i,z_j)-\ln 10/5\right]^2}{\sum_{i,j}\mathcal{C}^{-1}_{SNIa}(z_i,z_j)} ,            \\
&&\mathcal{B}\equiv -2\ln\left( \frac{\ln 10}{5}\sqrt{\frac{2\pi}{\sum_{i,j}\mathcal{C}^{-1}_{SNIa}(z_i,z_j)}}\right).
\end{eqnarray}

We also take into account Gamma Ray Bursts (GRBs) proposed as most luminous astrophysical objects at high redshift as  the complementary standard candles.  For GRBs, the $\chi^2_{GRBs}$ is given by:
\begin{eqnarray}
\chi^2_{GRBs}=\sum_{i}\frac{\mathcal{M}^2(z_i;\{\Theta_p\})}{\sigma_i^2}+\mathcal{A}_{GRBs}+\mathcal{B}_{GRBs},
\end{eqnarray}
where
\begin{eqnarray}
\mathcal{A}_{GRBs}&\equiv&-\frac{\left[\sum_{i}\frac{\mathcal{M}(z_i;\{\Theta_p\})}{\sigma_i^2}-\ln 10/5\right]^2}{\sum_{i}\frac{1}{\sigma_i^2}},             \\
\mathcal{B}_{GRBs}&\equiv&-2\ln\left( \frac{\ln 10}{5}\sqrt{\frac{2\pi}{\sum_{i}\frac{1}{\sigma_i^2}}}\right).
\end{eqnarray}
Finally for SNIa and GRBs observations we construct $\chi^2_{SG}\equiv\chi^2_{SNIa}+\chi^2_{GRBs}$.
In this paper we used recent Joint Light-curve Analysis (JLA) sample constructed from the SNLS and SDSS SNIa data, together with several samples of low redshift SNIa \cite{Ade:2015xua}. We also utilize the "Hymnium" sample including 59 samples for GRBs data set. These data sets have been extracted out of 109 long GRBs 
\cite{Wei:2010wu}.


\subsection{Baryon Acoustic Oscillations}
\begin{table}[ht]
\centering
\begin{tabular}{llcccc}
\hline
\hline Redshift & Data Set & $r_s/D_V(z;\{\Theta_p\})$ & Ref.\\ \hline
0.10   &  6dFGS     & $0.336\pm0.015$       &  \cite{Beutler:2011hx} \\
0.35 &  SDSS-DR7-rec    & $0.113\pm0.002$& \cite{Padmanabhan:2012hf} \\
0.57 &  SDSS-DR9-rec    & $0.073\pm0.001$ & \cite{Anderson:2012sa} \\
0.44 &  WiggleZ & $0.0916\pm0.0071$     &  \cite{Blake:2012pj} \\
0.60 &  WiggleZ & $0.0726\pm0.0034$     &  \cite{Blake:2012pj} \\
0.73 &  WiggleZ & $0.0592\pm0.0032$     &  \cite{Blake:2012pj} \\
\hline
\hline
\end{tabular}
\caption{\label{baodata} Observed data for BAO \cite{Hinshaw:2012aka}. }
\end{table}
Baryon Acoustic Oscillations or in brief BAO at recombination era are the footprint of oscillations in the baryon-photon plasma on the matter power spectrum. It can be utilized as a typical standard ruler, calibrated to the sound horizon 
at the end of the drag epoch. Since the acoustic scale is so large, BAO are largely unaffected by nonlinear evolution. The BAO data can be applied to measure both the angular diameter distance $D_A(z;\{\Theta\})$, and the expansion rate of the Universe $H(z;\{\Theta\})$ either separately or through their combination as \cite{Ade:2015xua}:
\begin{equation}
D_V(z;\{\Theta_p\}) = \left[ (1+z)^2 D_A^2 (z;\{\Theta\}) \frac{cz}{H(z;\{\Theta_p\})} \right]^{1/3},
\end{equation}
where $D_V (z;\{\Theta_p\})$ is volume-distance. The distance ratio used as BAO criterion is defined by:
\begin{eqnarray}
d_{BAO}(z;\{\Theta_p\})\equiv \frac{r_{s}(z;\{\Theta_p\})}{D_V(z;\{\Theta_p\})},
\end{eqnarray}
here $r_{s}(z;\{\Theta_p\})$ is the comoving sound horizon.  In this paper to take into account different aspects of BAO observations and improving our constraints, we use 6 reliable measurements of BAO indicators including Sloan Digital Sky Survey (SDSS) data release 7 (DR7) \cite{Padmanabhan:2012hf},  SDSS-III Baryon Oscillation Spectroscopic Survey (BOSS) \cite{Anderson:2012sa}, WiggleZ survey \cite{Blake:2012pj} and 6dFGS survey \cite{Beutler:2011hx}. BAO observations contain 6 measurements from redshift interval, $z\in [0.1,0.7]$ (for observed values at higher redshift one can refer to \cite{Delubac:2014aqe}). The observed values for mentioned redshift interval have been reported in Tab. \ref{baodata}. Also the inverse of covariance matrix is given by  
\begin{widetext}
\begin{eqnarray}\label{covbao}
{\mathcal C}^{-1}_{BAO} = \left(\begin{array}{rrrrrr}
4444.4 & 0 & 0 & 0 & 0 & 0 \\
0 & 34.602 & 0 & 0 & 0 & 0 \\
0 & 0 & 20.661157 & 0 & 0 & 0 \\
0 & 0 & 0 & 24532.1  & -25137.7 & 12099.1 \\
0 & 0 & 0 & -25137.7 & 134598.4 & -64783.9 \\
0 & 0 & 0 & 12099.1 & -64783.9 & 128837.6 \\
\end{array}
\right).
\end{eqnarray}
\end{widetext}
Therefore $\chi^2_{BAO}$ is written by:
\begin{eqnarray}
\chi^2_{BAO}\equiv \Delta d^{\dag}\cdot {\mathcal C}^{-1}_{BAO}\cdot \Delta d.
\end{eqnarray}
In above equation $\Delta d(z;\{\Theta_p\})\equiv d_{obs}(z)-d_{BAO}(z;\{\Theta_p\})$ and ${\mathcal C}^{-1}_{BAO}$ is given by Eq. (\ref{covbao}). The $d_{obs}(z)$ is reported in Tab. \ref{baodata}.

\subsection{CMB observations}
Another part of data to put observational constraints on free parameter of viscous dark energy model is devoted to CMB observations. 
Here we use following likelihood function for CMB power spectrum observations:
\begin{eqnarray}
\chi^2_{CMB-power}=\Delta C^{\dag}\cdot \mathcal {M}_{CMB}^{-1}\cdot \Delta C,
\end{eqnarray}
here $\Delta C_{\ell}\equiv C_{\ell}^{obs}-C_{\ell}(\{\Theta_p\})$ and $\mathcal {M}_{CMB}$ is covariance matrix for CMB power spectrum. As a complementary part for CMB observational constraints, we also used CMB lensing from SMICA pipeline of {\it Planck} 2015. To compute CMB power spectrum for our model, we used Boltzmann code CAMB \cite{Lewis:1999bs}. Here, we don't consider  dark energy clustering, consequently, at the first step, perturbations in radiation, baryonic and cold dark matters are mainly affected by background evolution which is modified due to presence of viscous dark energy. However the precise computation of perturbations should be taken into account perturbation in viscous dark energy, but since DE in our model at the early universe is almost similar to cosmological constant, consequently mentioned terms at this level of perturbation can be negligible. This case is out of the scope of current paper and we postpone such accurate considerations for an other study. We made careful modification to CAMB code and combined with publicly available cosmological  Markov Chain Monte Carlo code CosmoMC  \cite{Lewis:2002ah}.

\subsection{HST-Key project}

In order to adjust better constraint on the local expansion rate of Universe, we use Hubble constant measurement from Hubble Space Telescope (HST). 
Therefore additional observational point for analysis is \cite{Riess:2011yx} : $$H_0=73.8 \pm 2.4 km s^{-1}Mpc^{-1}.$$

In the next section, we will show the results of our analysis for best fit values for free parameters and their confidence intervals for one and two dimensions.

\section{Results and discussion}
As discussed in previous section, to examine the consistency of our viscous dark energy model with most recent observations, we use following tests:
\begin{enumerate}
\item SNIa luminosity distance from Joint Light-curve Analysis (JLA) which made from SNLS and SDSS SNIa compilation.
\item GRBs data sets for large interval of redshift as a complementary part for luminosity distance constraints.
\item BAO data from galaxy surveys  SDSS DR11, SDSS DR11 CMASS, 6dF.
\item CMB temperature fluctuations angular power spectrum from {\it Planck} 2015 results.
\item CMB lensing from SMICA pipeline of {\it Planck} 2015.
\item Hubble constant measurement from Hubble Space Telescope (HST) ($H_0=73.8 \pm 2.4 km s^{-1}Mpc^{-1}$) with flat prior.
\item Cosmographic distance ratio test.
\item Hubble parameter for different redshifts.
\item Cosmic age test.

\end{enumerate}

\begin{table}[H]
\centering
\begin{tabular}{llccc}
\hline
\hline Parameter & JLA & JLA+GRBs  \\
\hline $\Omega_b h^2 $ &$0.0232^{+0.0019}_{-0.0027}$&$0.0233^{+0.011}_{-0.0059} $ \\
\hline $\Omega_{\rm m}h^2$ &$0.1175^{+0.0073}_{-0.0089}$ &$0.112^{+0.022}_{-0.022}$   \\
\hline $\gamma$ &$0.1386^{+0.0034}_{-0.0024}$&$0.126\pm 0.043$   \\
\hline $\Omega_{\rm DE}^0$ &  $0.701\pm 0.025$&$0.702\pm 0.033$  \\
\hline $H_0 $ & $68.8^{+2.1}_{-2.8} $&$71.5^{+6.8}_{-6.7}$\\
\hline
\hline
\end{tabular}
\caption{\label{tbl:jla} Best fit values for viscous dark energy using JLA and combination with GRBs at $68 \%$ confidence interval. }
\end{table}

\begin{table}[]
\centering
\begin{tabular}{llcccc}
\hline
\hline Parameter & BAO & HST & JGBH \\
\hline $\Omega_b h^2 $ & $0.02277^{+0.00092}_{-0.0017}$&$0.0222^{+0.0022}_{-0.0016} $ & $0.0223^{+0.0013}_{-0.0021}$\\
\hline $\Omega_{\rm m} h^2$ &$0.1194^{+0.0031}_{-0.0044}$&$0.111^{+0.011}_{-0.0088}$ &$0.1181^{+0.0018}_{-0.0022}$  \\
\hline $\gamma$ &$0.1406^{+0.0032}_{-0.0024}$&$0.1403^{+0.0014}_{-0.0012}$ & $0.1404\pm 0.0014$  \\
\hline $\Omega_{\rm DE}^0 $ & $0.692\pm 0.012$&$0.734^{+0.031}_{-0.048}$&$0.696\pm 0.010$   \\
\hline $H_0 $ &$68.07^{+0.85}_{-1.3}$&$71.4^{+2.6}_{-3.6}$&$68.1\pm 1.3$ \\
\hline
\hline
\end{tabular}
\caption{\label{tbl:bestvalues1} Bayesian $68 \%$ confidence limits for model based on JLA, BAO, HST  \& JLA+GRBs+BAO+HST. }
\end{table}

 \begin{table}[]
\centering
\begin{tabular}{llcccc}
\hline
\hline Parameter & {\it Planck} TT & CMB Lensing & {\it Planck} TT +JGBH\\
\hline $\Omega_b h^2 $ &$0.02210^{+0.00037}_{-0.00036}$ &$0.022^{+0.021}_{-0.012}   $ & $0.02220^{+0.00033}_{-0.00032}$\\
\hline $\Omega_m h^2$ &$0.1198^{+0.0045}_{-0.0044}$ & $0.118^{+0.036}_{-0.040}   $ &$0.1182^{+0.0025}_{-0.0024}$ \\
\hline $\gamma$ &$0.32^{+0.31}_{-0.26}$& $0.143^{+0.026}_{-0.016}   $ & $0.26^{+0.22}_{-0.20}$ \\
\hline $\Omega_{\rm DE}^0 $ &  $0.684^{+0.026}_{-0.028}$ &  $0.69^{+0.18}_{-0.12} $ & $0.694^{+0.014}_{-0.015}$\\
\hline $H_0 $ & $67.2^{+1.9}_{-1.9}$ & $69.0^{+2.0}_{-1.0} $ & $67.9\pm1.1$\\

\hline
\hline
\end{tabular}
\caption{\label{tbl:bestvalues2} Bayesian $68 \%$ confidence limits for model based on {\it Planck} TT, CMB Lensing and {\it Planck} TT+JGBH. }
\end{table}
\begin{figure}
\centering
       \includegraphics[width=.9\columnwidth]{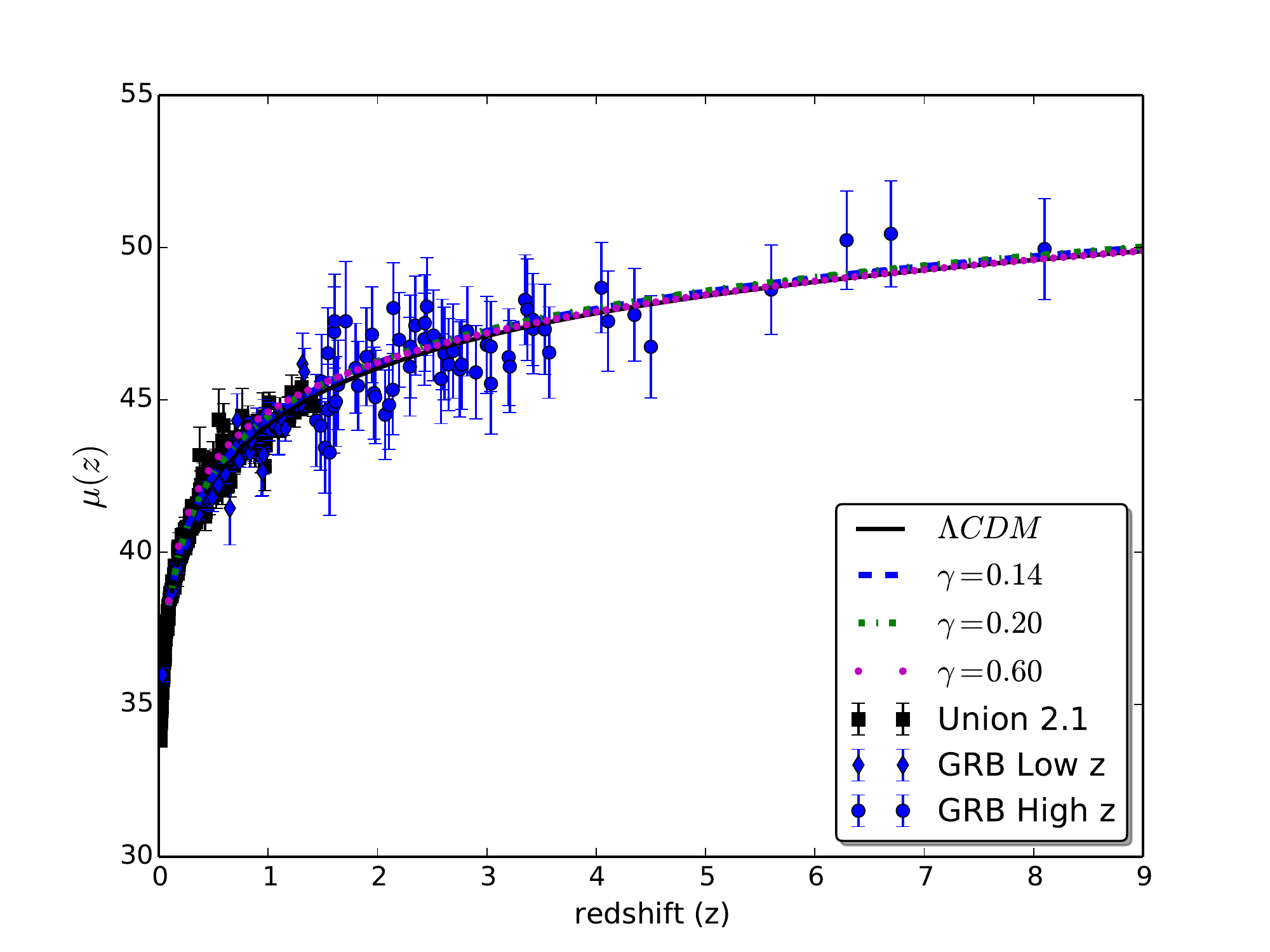}\\
        \includegraphics[width=.9\columnwidth]{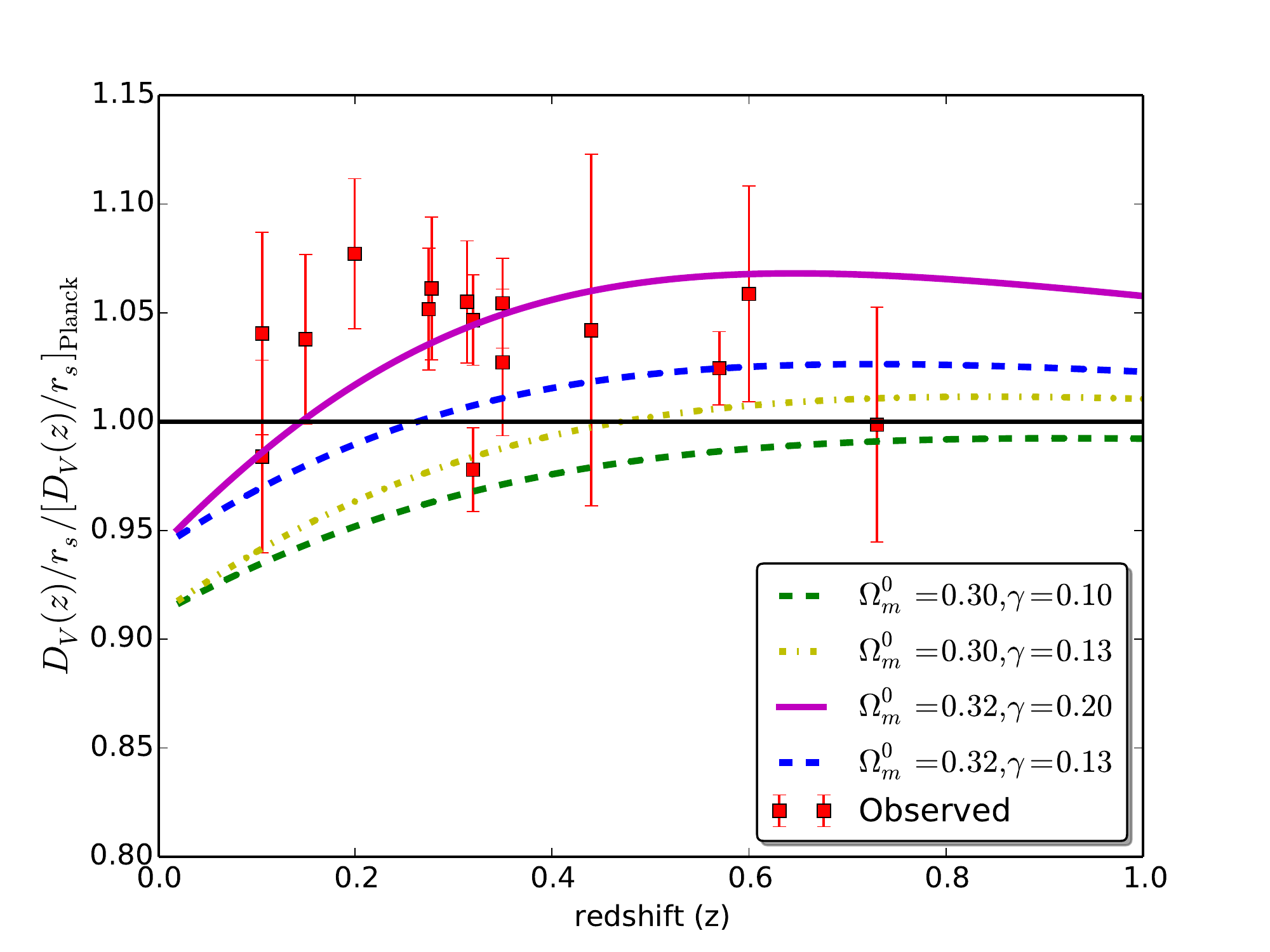}
    \caption{\textit{Upper panel:}  Distance modulus of bulk viscous model compared with JLA and GRBs data. \textit{Lower panel:} BAO observables for viscous dark energy and $\Lambda$CDM models. The observed data sets is reported in Table \ref{baodata}. }
\label{fig:luminosity-dist-modul-dist}
\end{figure}

In the upper panel of Fig. \ref{fig:luminosity-dist-modul-dist}, we plotted luminosity distance for $\Lambda$CDM best fit (solid line) and viscous dark energy model for different values of $\gamma$. In our viscous dark energy model, the EoS at late time belongs to Phantom type, therefore by increasing the value of $\gamma$ when other parameters are fixed, the contribution of viscous dark energy becomes lower than cosmological constant.  Subsequently,  the distance modulus is longer than that of for cosmological constant. For complementary analysis, we also added GRBs results for observational constraint. Lower panel indicates BAO observable quantity. Higher value of viscosity for dark energy model causes to good agreement between model and observations. In such case, the consistency with early observations decreases, therefore, there is a trade off in determining $\gamma$ with respect to late and early time observational data sets. Marginalized posterior probability function for various free parameters of viscous dark energy model have been indicated in Fig. \ref{fig:likelihood-prob}. In this figure we find that different observations to confine, $\gamma$, $\Omega^0_{\rm{DE}}$ and $H_0$ are almost consistent. The results of Bayesian analysis to find best fit values for free parameters of viscous dark energy model (Table \ref{prior}) at $1\sigma$ confidence interval have been reported in Table \ref{tbl:jla}. In Table \ref{tbl:bestvalues1} the best fit values for parameters using BAO, HST and the combination of observations,  JLA$+$GRBs$+$BAO$+$HST (JGBH) have been reported at $68\%$ confidence interval.

\begin{figure}
\centering
  \includegraphics[width=.9\columnwidth]{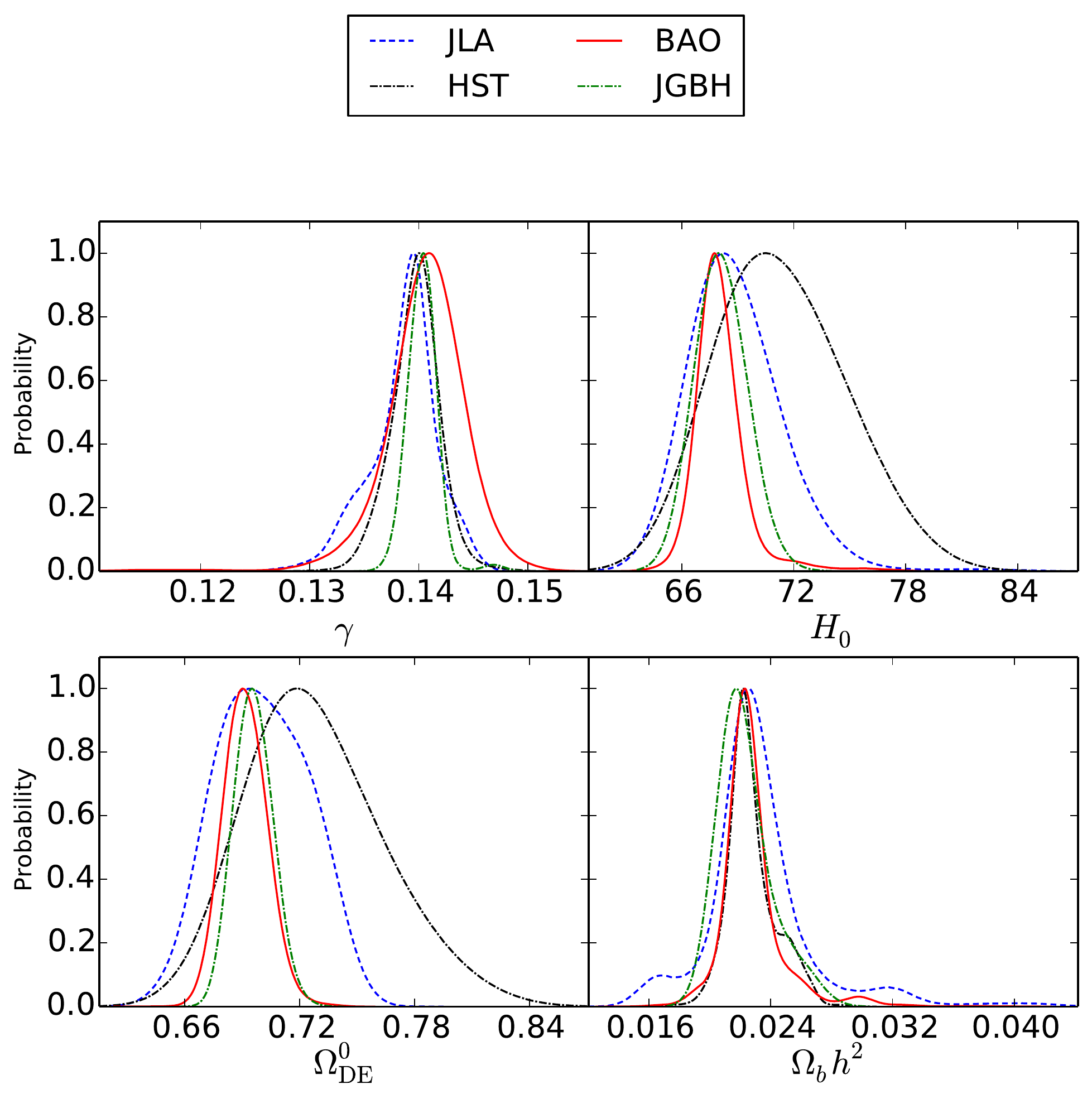}
  \caption{Marginalized posterior function for various free parameters of dynamical dark energy model. Dashed line represents observational constraint by joint analysis JLA$+$GRBs. Solid line corresponds to BAO analysis while long-dashed line indicates observational consistency for HST project.}
  \label{fig:likelihood-prob}
  \end{figure}

\begin{figure}
\centering
  \includegraphics[width=.9\columnwidth]{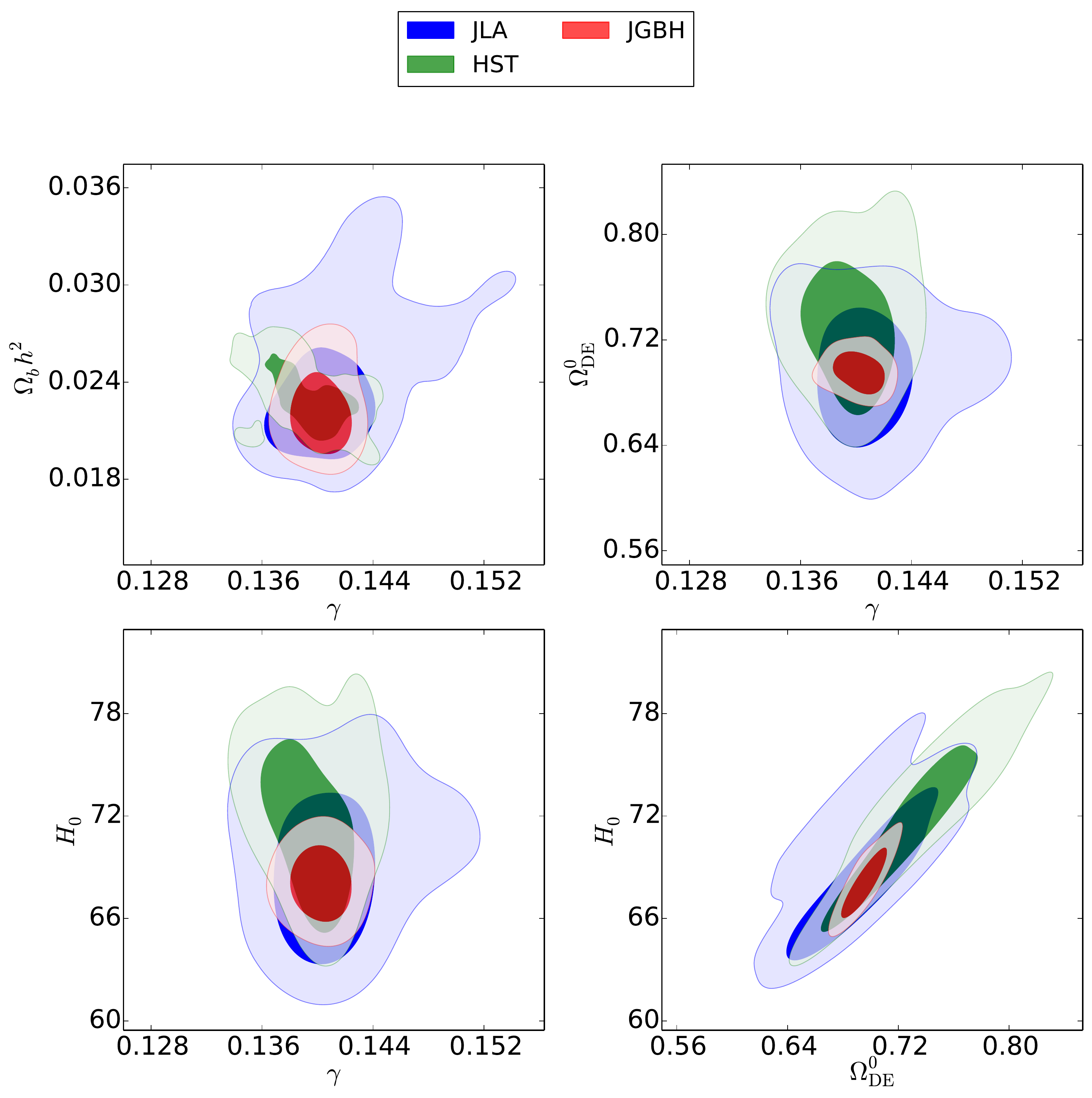}
  \caption{ Marginalized confidence regions at $68 \%$ and $95 \%$ confidence levels.}
  \label{fig:contours}
  \end{figure}

\begin{figure}
\centering
  \includegraphics[width=.9\columnwidth]{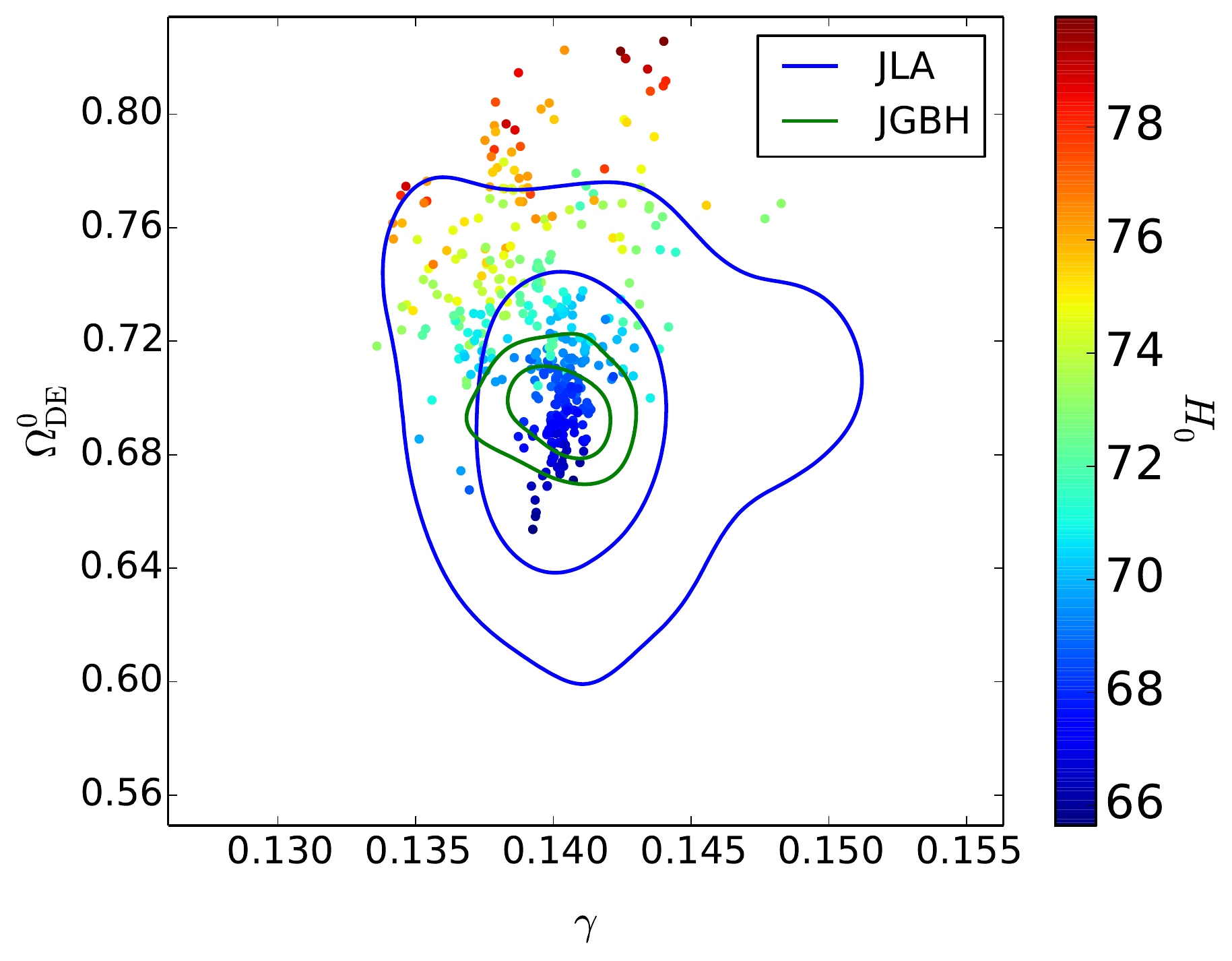}
  \caption{The effect of varying $H_0$ on the degeneracy between $\Omega_{\rm DE}^0-\gamma$ in the contour enclosing $68\%$ and $95\%$ confidence intervals.}
  \label{fig:3D-gama-omegal-H02}
  \end{figure}

  \begin{figure}
\centering
  \includegraphics[width=.9\columnwidth]{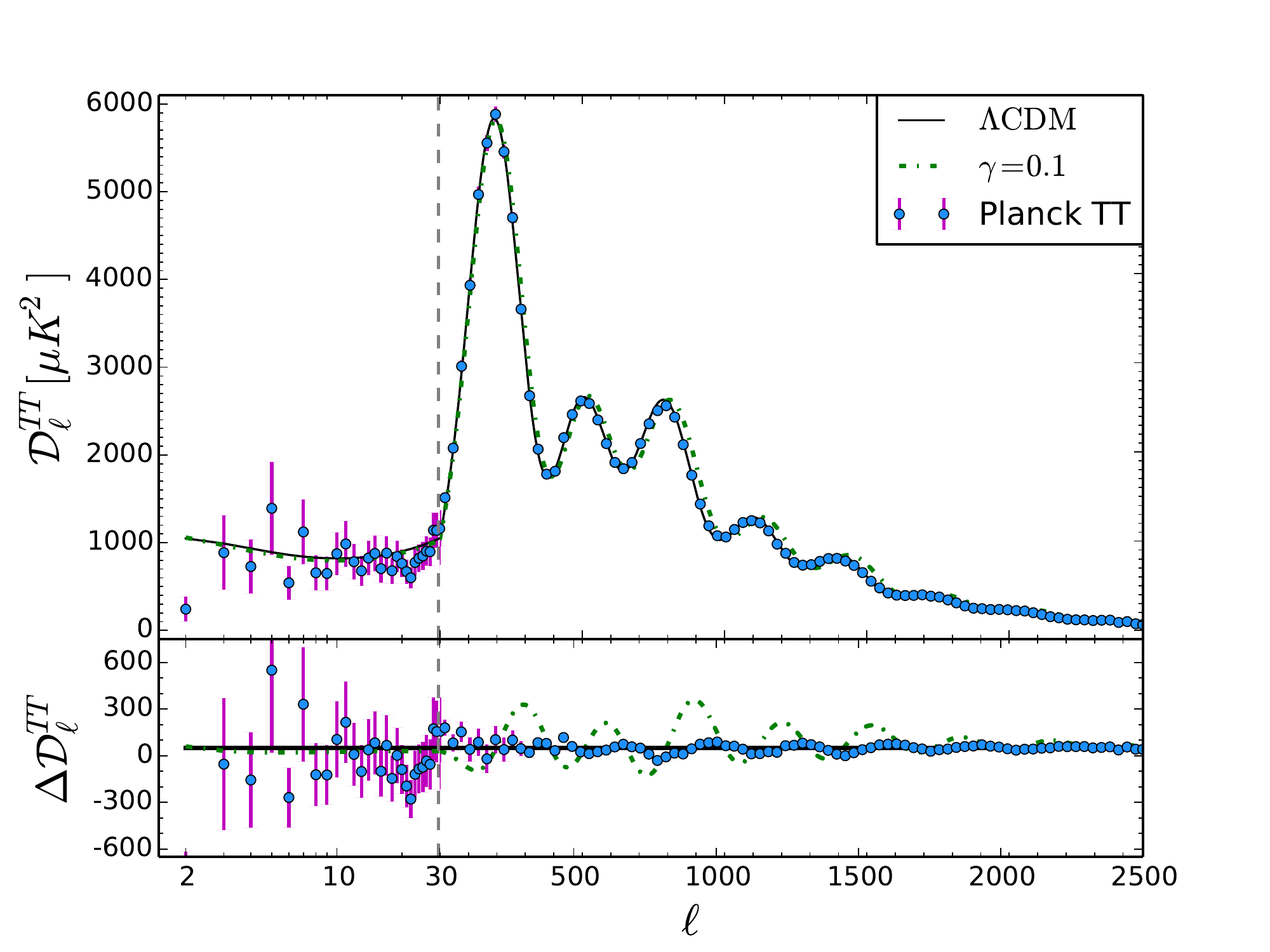}\\
  \includegraphics[width=.9\columnwidth]{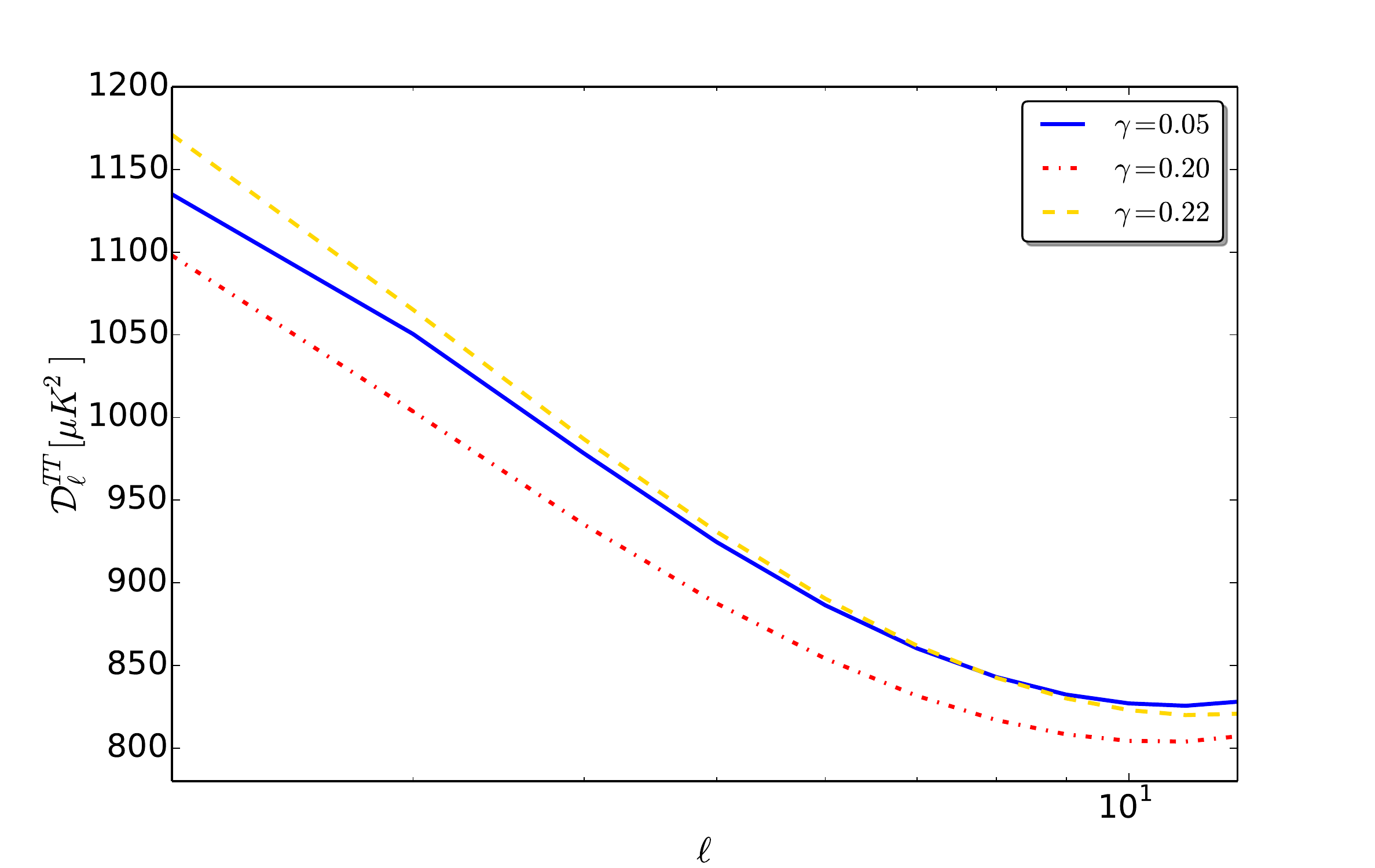}
  \caption{\emph{Upper panel}: CMB power spectrum for bulk viscous and $\Lambda$CDM model. \emph{Lower panel}: The viscous coefficient leads to ups and downs in power spectrum for higher $\ell$ while for small $\ell$ due to ISW contribution the plateau of power varies depending on value of $\gamma$. }
  \label{fig:powerspectrum}
 \end{figure}

The contour plots for various pairs of free parameters are indicated in  Fig. \ref{fig:contours}. Our results demonstrate that there exists acceptable consistency between different observations in determining best fit values for free parameters of viscous dark energy model.  Fig. \ref{fig:3D-gama-omegal-H02} illustrates the contour plot in the $\Omega_{\rm DE}^0-\gamma$ plane. The effect of changing the Hubble constant at the present time on the degeneracy of mentioned parameters  representing that by increasing the value of $H_0$, the best fit value for viscous dark energy density at present time increases while the viscosity parameter is almost not sensitive.

It turns out that at the early Universe the cosmological constant
has no role in evolution of the Universe,  on the contrary,  it is
an opportunity for a dynamical  dark energy  to contribute
effectively on that epoch. However according to Fig.
\ref{fig:density-ratio-a} the viscous dark energy to cold dark
matter ratio at early epoch asymptotically goes to zero,
nevertheless, we use CMB observation to examine the consistency of
our viscous model. In Fig. \ref{fig:powerspectrum}, we indicate the
behavior of our dynamical dark energy model on the power spectrum of
CMB. The higher value of $\gamma$, the lower contribution of
dynamical    dark energy model at the late time (see Fig.
\ref{fig:density-ratio-a}) resulting in lower value of ISW effect.
Consequently, the value of $\ell(\ell+1)C_{\ell}$ for small $\ell$
goes down. For $\gamma>\gamma_{\times}$ due to changing the nature
of  dark energy model, mentioned behavior is no longer valid and the
Sachs-Wolfe plateau becomes larger (see lower panel of
Fig. \ref{fig:powerspectrum}).
Considering {\it Planck} TT data, the best fit values are
$\gamma=0.32^{+0.31}_{-0.26}$ and $\Omega_{\rm DE}^0=0.684^{+0.026}_{-0.028}$ at $68\%$ confidence interval in flat Universe. {\it Planck} TT observation does't provide strong constraint on viscous parameter, $\gamma$. 
CMB lensing observational data puts strict constraint on $\gamma$ coefficient as reported in Table \ref {tbl:bestvalues2}.

Combining JLA$+$GRBs$+$BAO$+$HST  with  {\it Planck} TT data  leads to tight constraint on both $\gamma$ and $\Omega_{\rm{DE}}^0$ 
On the other hand, if we combine  JLA, GRBs, BAO and HST (JGBH), our results demonstrate that $\gamma=0.1404\pm 0.0014$ and $\Omega_{\rm DE}^0=0.696\pm0.010$ consequently accuracy improves remarkably.

 Tension in value of $H_0$ in this model is disappeared if we compare the best fit value for $H_0$ using local observations and that of determined by CMB.
This result is due to the early behavior of our viscous dark energy model.

\begin{figure}
\centering
  \includegraphics[width=.9\columnwidth]{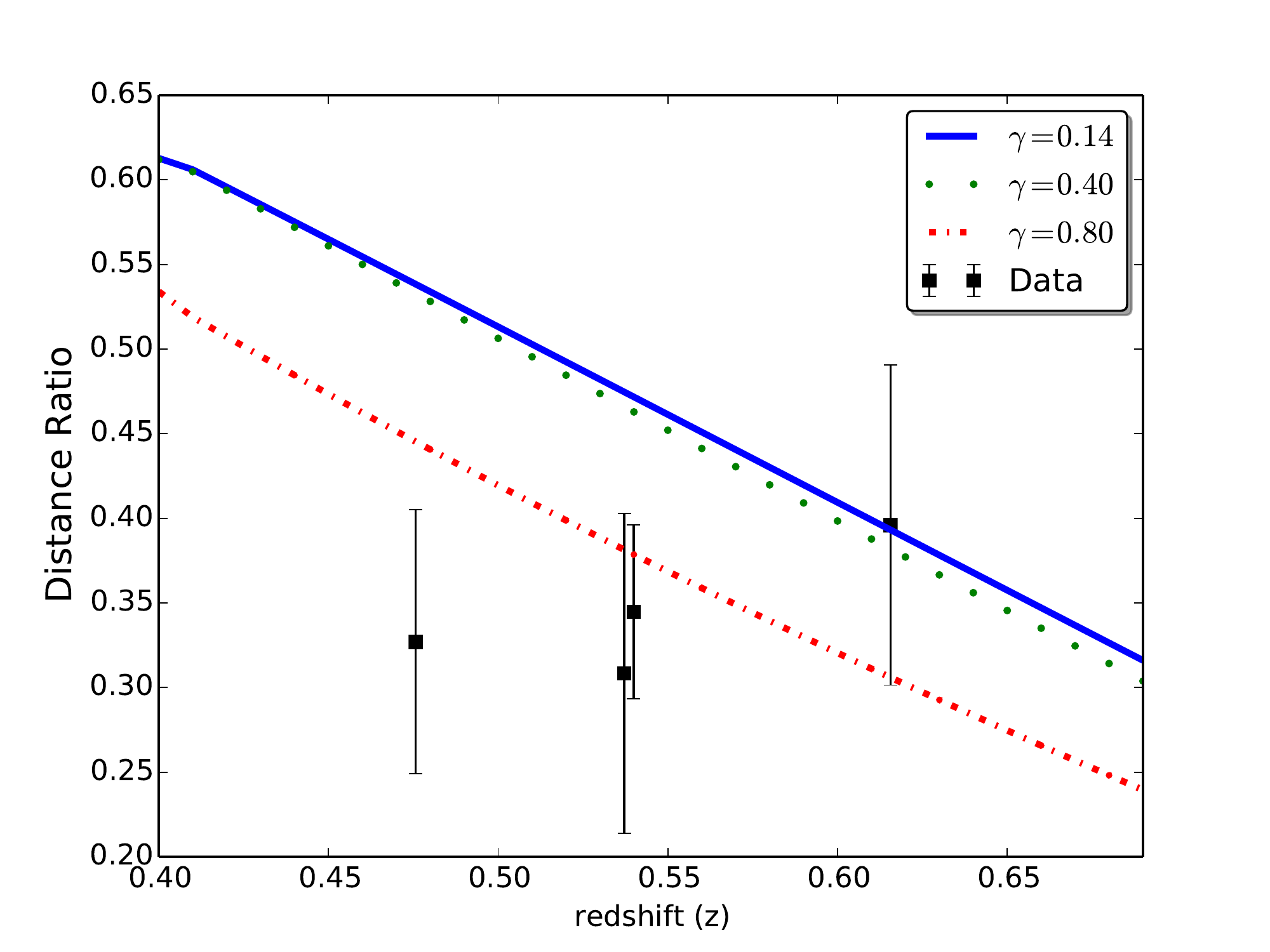}
    \caption{Cosmographic distance ratio for viscous dark energy model. Solid line corresponds to $\gamma=0.14$ according to best fit by JLA observation, dotted lines indicates  for $\gamma=0.40$ and dash-dot is for $\gamma=0.80$. Observed data are given from \cite{Miyatake:2016gdc}.}
  \label{fig:cgdr}
\end{figure}

In what follows we deal with  cosmographic distance ratio, Hubble parameter and cosmic age to examine additional aspect of viscous dark energy model.  Recently H. Miyatake et al., used the cross-correlation optical weak lensing and CMB lensing and introduced purely geometric quantity. This quantity is so-called cosmographic distance ratio defined by \cite{Jain:2003tba,Taylor:2006aw,Kitching:2008vx}:
\begin{equation}
r\equiv \frac{D_A(a_{\rm L},a_{\rm g})D_A(a_{\rm c})}{D_A(a_{\rm L},a_{\rm c})D_A(a_{\rm g})},
\end{equation}
where $D_A$ is angular diameter distance. $a_{\rm L}$, $a_{\rm g}$ and $a_{\rm c}$ are scale factors for lensing structure, the background galaxy source plane and CMB, respectively.
Here we used three observational values reported in  \cite{Miyatake:2016gdc} based on CMB and Galaxy lensing.
 Fig. \ref{fig:cgdr} represents $r$ as a function of redshift for the best values constrained by various observations in viscous  dark energy model.  Higher value of viscosity leads to better coincidence with current observations. Since in our model the maximum best value for viscosity is given by {\it Planck} TT observation which is  $\gamma=0.32^{+0.31}_{-0.26}$ at $1\sigma$ confidence interval, so one can conclude that  current observational data for $r$ have been mostly affected by CMB when we consider viscous dark energy model as a dynamical dark energy.

\begin{table}
\caption{ \label{tbl:hubble-data1} $H(z)$ measurements (in unit [$\mathrm{km\,s^{-1}Mpc^{-1}}$]) and their errors \cite{Farooq:2013hq}.}
\begin{center}
\label{hubble}
\begin{tabular}{ccc}
\hline\hline
~$z$ & ~~$H(z)$ &~~ $\sigma_{H}$\\
0.070&~~    69&~~~~~~~  19.6\\
0.100&~~    69&~~~~~~~  12\\
0.120&~~    68.6&~~~~~~~    26.2\\
0.170&~~    83&~~~~~~~  8\\
0.179&~~    75&~~~~~~~  4\\
0.199&~~    75&~~~~~~~  5\\
0.200&~~    72.9&~~~~~~~    29.6\\
0.270&~~    77&~~~~~~~  14\\
0.280&~~    88.8&~~~~~~~    36.6\\
0.350&~~    76.3&~~~~~~~    5.6\\
0.352&~~    83&~~~~~~~  14\\
0.400&~~    95&~~~~~~~  17\\
0.440&~~    82.6&~~~~~~~    7.8\\
0.480&~~    97&~~~~~~~  62\\
0.593&~~    104&~~~~~~~ 13\\
0.600&~~    87.9&~~~~~~~    6.1\\
0.680&~~    92&~~~~~~~  8\\
0.730&~~    97.3&~~~~~~~    7.0\\
0.781&~~    105&~~~~~~~ 12\\
0.875&~~    125&~~~~~~~ 17\\
0.880&~~    90&~~~~~~~  40\\
0.900&~~    117&~~~~~~~ 23\\
1.037&~~    154&~~~~~~~ 20\\
1.300&~~    168&~~~~~~~ 17\\
1.430&~~    177&~~~~~~~ 18\\
1.530&~~    140&~~~~~~~ 14\\
1.750&~~    202&~~~~~~~ 40\\
2.300&~~    224&~~~~~~~ 8\\

\hline\hline
\end{tabular}
\end{center}
\end{table}

We inspect the Hubble parameter in our model. To this end, by using Eq. (\ref{eq:hubble}) one can compute the rate of  expansion as a function if redshift and compare it with observed Hubble parameter  listed in Table. \ref{tbl:hubble-data1} for various redshift \cite{Farooq:2013hq} .

Fig. \ref{fig:hubblerate} shows Hubble expansion rate for viscous dark energy model.  There is a good agreement between this model and observed values for small $\gamma$. In Table. \ref{tbl:hubble-data1} we summarize all $H(z)$ measurements  used in upper panel of Fig. \ref{fig:hubblerate}.
 One can rewrite $\mathcal{H}(z;\{\Theta\})\equiv H(z;\{\Theta\})/H_0$ and we introduce
 the relative difference $\Delta \mathcal{H}(z;\{\Theta\})$  as follows:
\begin{equation}
\Delta \mathcal{H}(z;\{\Theta\})=100 \times \left[ \frac{\mathcal{H}(z;\{\Theta\})}{\mathcal{H}_{\Lambda CDM}(z)} -1 \right].
\end{equation}
Above quantity has been illustrated in lower panel of Fig. \ref{fig:hubblerate} as a function of redshift. For higher value of $\gamma$ at small redshift, we get pronounce difference between viscous dark energy and cosmological constant.

\begin{figure}
\centering
  \includegraphics[width=.9\columnwidth]{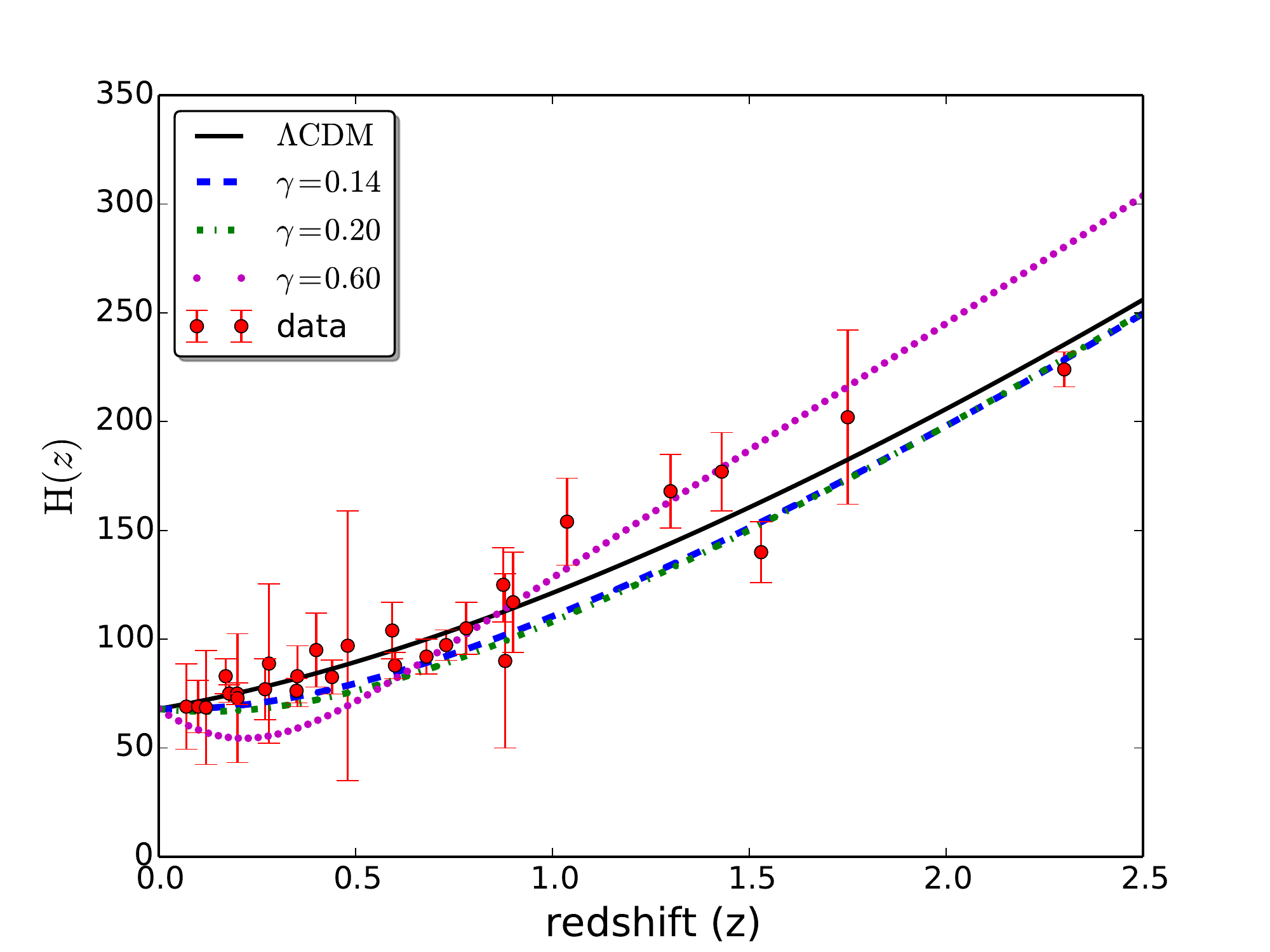}\\
  \includegraphics[width=.9\columnwidth]{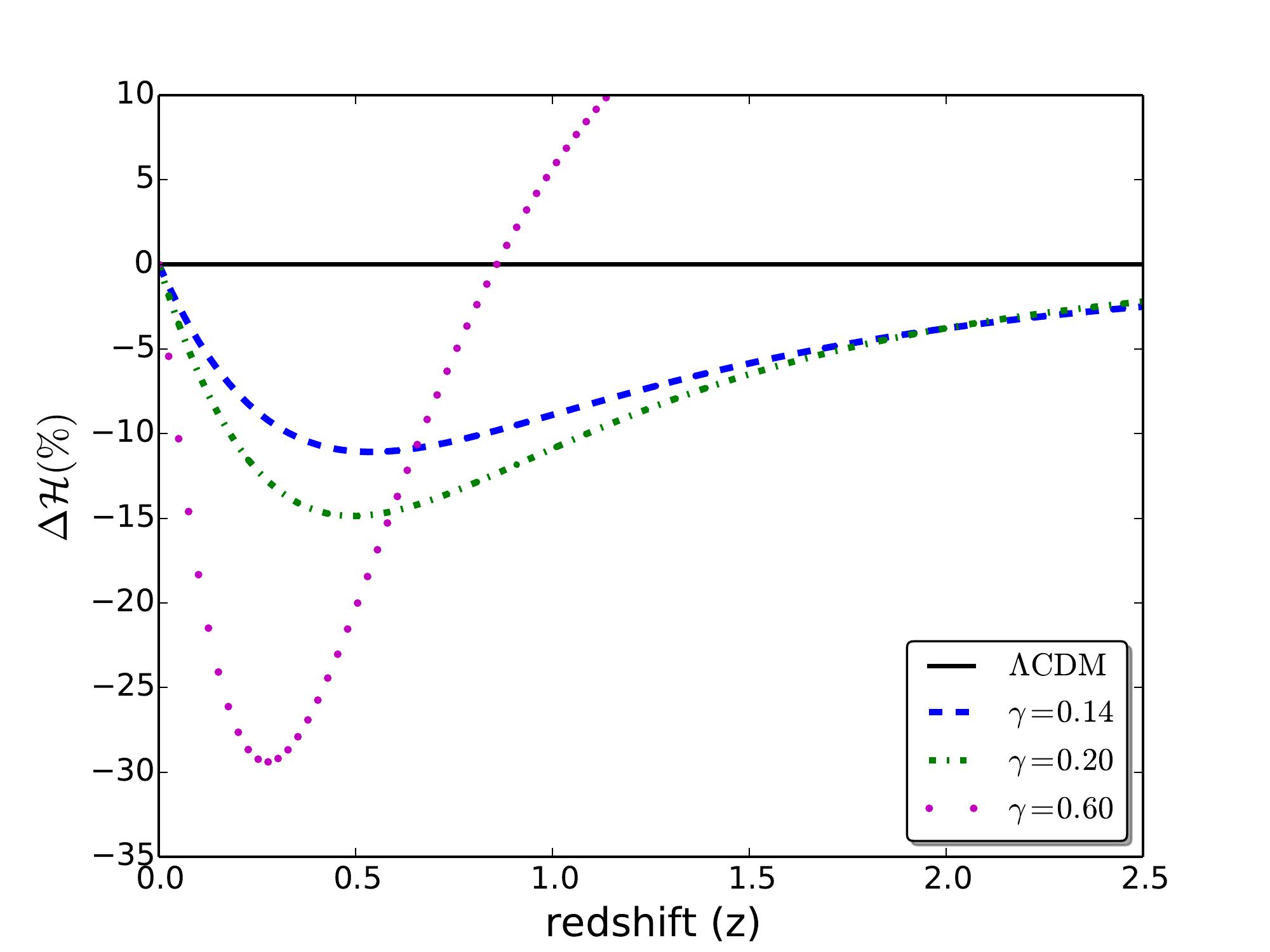}
    \caption{\textit{Upper panel:} Hubble parameter in comparison with various observational values for $H(z)$. \textit{Lower panel:} relative difference of Hubble parameter as a function of redshift for various values of viscosity. }
  \label{fig:hubblerate}
\end{figure}

The cosmic age crisis is long standing subject in the cosmology and is a proper litmus test to examine dynamical dark energy model. Due to many objects observed at intermediate and high redshifts, there exist some challenges to accommodate some of Old High Redshift Galaxies (OHRG) in  $\Lambda$CDM model \cite{Jain:2005gu}. There are many suggestion to resolve mentioned cosmic age crisis \cite{Tammann:2001eg,Cui:2010dr,Rahvar:2006tm}, however this discrepancy has not completely removed yet and it becomes as {\it smoking-gun} of evidence for the advocating  dark energy component. As an illustration, Simon et.al. demonstrated the {\it lookback time redshift} data by computing the age of some old passive galaxies at the redshift interval $0.11\le z\le1.84$ with 2 high redshift radio galaxies at $z=1.55$ which is named as LBDS $53$W$091$ with age equates to $3.5$-Gyr \cite{Dunlop:1996mp,Spinrad:1997md} and the LBDS $53$W$069$ a $4.0$-Gyr  at $z=1.43$ \cite{Dunlop1999}, totally they used 32 objects  \cite{Simon:2004tf}.
In addition, 9 extremely old globular clusters older than the present cosmic age based on 7-year WMAP observations have been found in  \cite{Wang:2010su}. Beside mentioned objects to check the {\it smoking-gun} of dark energy models,  a quasar APM $08279+5255$ at $z=3.91$,  with an age equal to $t=2.1_{-0.1}^{+0.9}$Gyr is considered \cite{Hasinger:2002wg}.

To check the age-consistency, we introduce a quantity as:
\begin{eqnarray}
 \tau(z_i;\{\Theta_p\})&=&\frac{t(z_i;\{\Theta_p\})}{t_{obs}(z_i)},  \quad i=1...42,
\end{eqnarray}
where $t(z_i;\{\Theta\})$ is the age of Universe computed by Eq. (\ref{age}) and
$t_{obs}(z_i)$ is an estimation for the age of $i$th old cosmological object. In above equation $\tau\ge 1$ corresponds to compatibility of model based on observed objects. 9 extremely old globular clusters  are located  in M31 and we report the $\tau$ value for them in Table \ref{tabagetau}. As indicated in mentioned table, only B050 is in tension over $4\sigma$ with current observations. In addition in Fig. \ref{fig:age-galaxy}, we illustrate the value of $\tau$ for the rest of data for various observational constraints. At $2\sigma$ confidence interval, there is no tension with cosmic age in viscous dark energy model even for very old high redshift quasar, $08279+5255$, at $z=3.91$. Therefore,   all old objects used in age crisis analysis can accommodate by considering viscous dark energy model.

Finally,  considering bulk viscous model and by combining different observations, one can  improve age crisis in the frame work of very old globular clusters and high redshift objects.

\begin{widetext}
\begin{center}
\begin{table}[H]
\centering
\begin{tabular}{|c|c|c|c|c|}
\hline
\hline
 Name & JLA$+$GRBs & BAO&JGBH&TT$+$JGBH  \\\hline
B024& $0.959^{+0.055}_{-0.053}$  &  $0.951^{+0.049}_{-0.053}$ & $0.991^{+0.049}_{-0.049}$&$0.960^{+0.048}_{-0.048}$   \\\hline
B050  & $0.914^{+0.032}_{-0.029} $&$0.906^{+0.022}_{-0.030}$&$0.944^{+0.019}_{-0.019}$&$0.915^{+0.019}_{-0.019}$\\\hline
B129 & $0.969^{+0.053}_{-0.051}$  & $0.960^{+0.047}_{-0.051}$& $1.001^{+0.047}_{-0.047}$&$0.969^{+0.045}_{-0.046}$\\\hline

B144D  & $1.019^{+0.074}_{-0.072}$ &$1.010^{+0.068}_{-0.072}$ & $1.052^{+0.070}_{-0.070} $&$1.019^{+0.068}_{-0.068}$\\\hline

B239  & $ 1.009^{+0.146}_{-0.145} $ &$1.000^{+0.142}_{-0.144} $& $1.042^{+0.147}_{-0.147}$   &$1.010^{+0.143}_{-0.143}$\\\hline
B260  &$1.023^{+0.047}_{-0.044} $   &$1.014^{+0.039}_{-0.045}$&$1.057^{+0.037}_{-0.038}$&$1.024^{+0.037}_{-0.037}$\\\hline
B297D  &$0.964^{+0.061}_{-0.059}$   &$0.955^{+0.055}_{-0.059}$&$0.995^{+0.056}_{-0.056}$&$0.964^{+0.054}_{-0.054}$\\\hline
B383   &$1.046^{+0.084}_{-0.083}$&$1.037^{+0.079}_{-0.082}$&$1.080^{+0.081}_{-0.081}$&$1.046^{+0.079}_{-0.079}$\\\hline
B495 &$1.006^{+0.049}_{-0.046}$&$0.997^{+0.041}_{-0.046}$&$1.039^{+0.040}_{-0.040}$&$1.007^{+0.039}_{-0.039}$\\\hline
\hline
\end{tabular}
\caption{\label{tabagetau} The $\tau$ value  for 9 old globular cluster located in M31 galaxy \cite{Wang:2010su}.}
\end{table}
\end{center}
\end{widetext}
\begin{figure}[ht]
\centering
  \includegraphics[width=1\columnwidth]{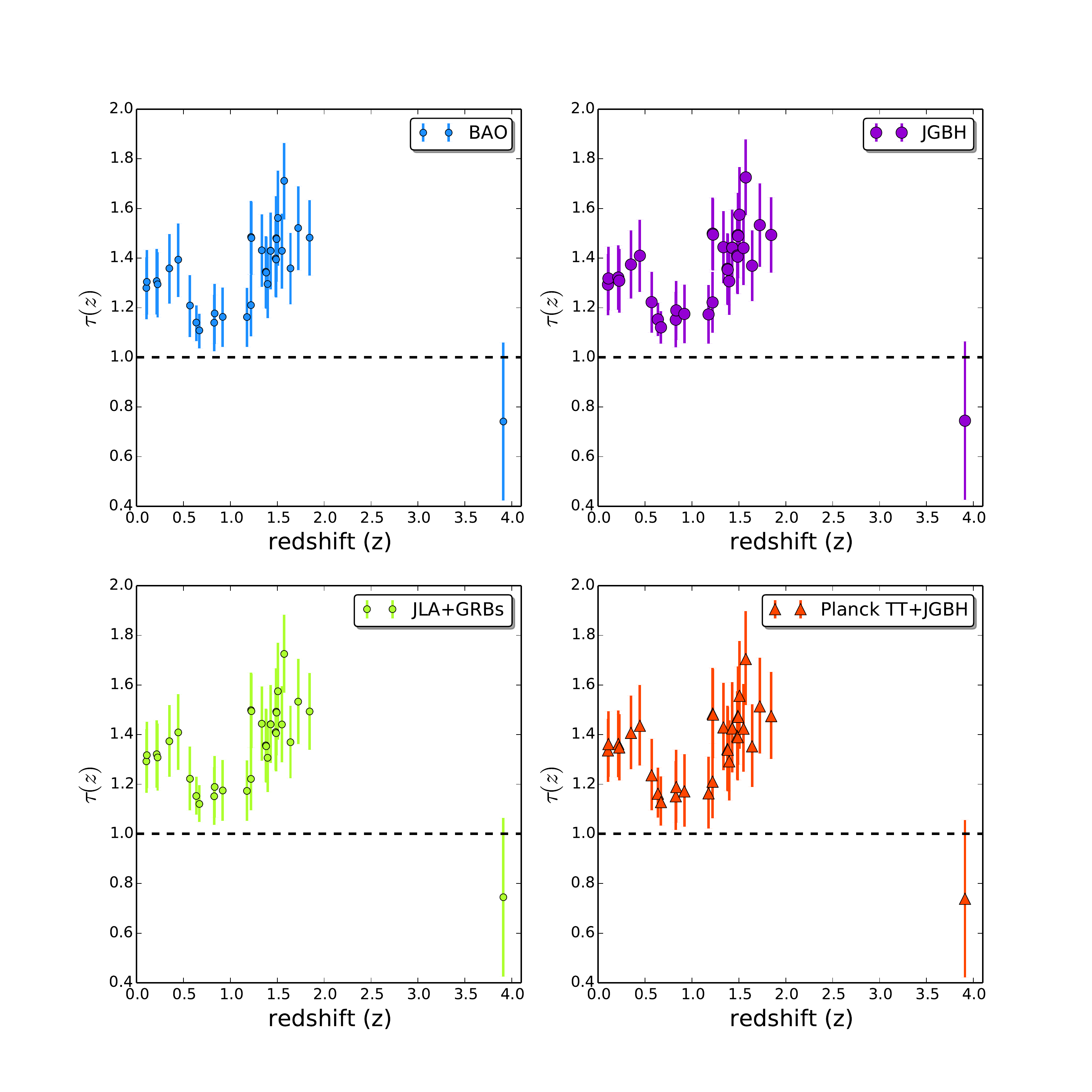}
    \caption{$\tau$ as a function of redshift for 32 old objects. Data have been given from \cite{Simon:2004tf,Dunlop1999,Hasinger:2002wg}.}
  \label{fig:age-galaxy}
\end{figure}

\section{Summary and Conclusions}\label{s:con}
In this paper, we examined a modified version for  dark energy model inspired by dissipative phenomena in fluids according to Eckart theory as the zero-order level for the thermodynamical dissipative process. In order to satisfy statistical isotropy, we assumed a special form for dark energy bulk viscosity. In this model, we have two components for energy contents without any interaction  between them. Our viscous dark energy model showed Phantom crossing avoiding Big-Rip singularity. Our results indicated that however the energy density of viscous dark energy becomes  zero at a typical scale factor, $\tilde{a}$ (Eq. (\ref{red-shift})), depending on viscous coefficient, interestingly there is no ambiguity for time definition in mentioned model (see Eq. (\ref{ageb}) for one component case).

We have also proposed a non-minimal derivative coupling scalar field with zero potential to describe viscous dark energy model for two components Universe. In this approach, coupling parameter is related to viscous coefficient and energy density of viscous dark energy at the present time. For zero value of $\gamma$, the standard action for canonical scalar field to be retrieved. To achieve real value for scalar field, $\epsilon$ should be negative. According to Eq. (\ref{kappap}), the coupling parameter is bounded according to $\kappa\in [-1/9H_0^2(1-\Omega_{\rm DE}^0), 0]$.  Evolution of $\tilde{\phi}$ indicated that scalar field has no monotonic behavior as scale factor increases. Subsequently, at $\tilde{a}$ the value of $\rho_{\phi}$ becomes zero corresponding to Phantom crossing era.

From observational consistency points of view, we examined the
effect of viscous dark energy model on geometrical parameters,
namely, comoving distance, Alcock-Paczynski test, comoving volume
element and age of Universe.  The comoving radius of Universe for
$\gamma<\gamma_{\times}$ has growing behavior indicating the Phantom
type of viscous dark energy. Alcock-Paczynski test showed that there is sharp
variation in relative behavior of viscous dark energy model with
respect to cosmological constant at low redshift. The redshift interval
for  occurring mentioned variation is almost independent from viscous
coefficient.  Comoving volume element increased by increasing the
viscous coefficient for $\gamma<\gamma_{\times}$ resulting in
growing number-count of cosmological objects.

To discriminate between different candidates for dark energy, there
are some useful criteria. In this paper,  cosmographic parameters
have been examined and relevant results showed that at the late time
it is possible to distinguish viscous dark energy from cosmological
constant. As indicated in Fig. \ref{fig:qz} at low redshift there is
a meaningful differences between bulk viscous model and $\Lambda
\rm{CDM}$ model. For completeness, we also used $Om$-diagnostic
method  to evaluate the behavior of viscous dark energy model. This
method is very sensitive to classify our model depending on $\gamma$
parameter. For small $\gamma$, $Om(z)$ represents the Phantom type
and it is possible to distinguish this model from $\Lambda$CDM.

To perform a systematic analysis and to put observational constraints on model free parameters reported in Table \ref{prior}, we considered supernovae, Gamma Ray Bursts, Baryonic acoustic oscillation, Hubble Space Telescope, {\it Planck} data for CMB observations. To compare  theoretical distance modulus with observations, we used recent SNIa catalogs and Gamma Ray Bursts including higher redshift objects. The best fit parameters according to SNIa by JLA catalog are: $\Omega_b h^2=0.0232^{+0.0019}_{-0.0027}$, $\Omega_{\rm DE}^0=0.701\pm0.025$, $\gamma=0.1386^{+0.0034}_{-0.0024}$ and $H_0=68.8^{+2.1}_{-2.8}$ at $68\%$ confidence limit. Doing joint analysis with GRBs leads to increase the viscous dark energy content at the present time. The best fit values from joint analysis of  JLA$+$GRBs$+$BAO$+$HST results in  $\Omega_{\rm DE}^0=0.696\pm0.010$, $\gamma=0.1404\pm 0.0014$ and $H_0=68.1\pm1.3$ at $1\sigma$ confidence interval (see Tabs. \ref{tbl:jla} and \ref{tbl:bestvalues1}).  According to {\it Planck} TT observation, the value of viscosity coefficient increases considerably and gets $\gamma=0.32^{+0.31}_{-0.26}$. Tension in Hubble parameter is almost resolved in the presence of viscous dark energy component. Joint analysis of JGBH$+${\it Planck} TT causing $H_0=67.9\pm 1.1$ at 68\% level of confidence. It is worth noting that using late time observations confirmed that viscous dark energy model at reliable level (see Fig. \ref{fig:likelihood-prob}) while according to our results reported in Tab. \ref{tbl:bestvalues2}, taking into account early observation such as CMB power spectrum, removes mentioned tight constraint. Marginalized contours have been illustrated in Fig. \ref{fig:contours}. The power spectrum of TT represented that, viscous dark energy model has small amplitude ups and downs for large $\ell$ which describes observational data better than $\Lambda$CDM.

As a complementary approach in our investigation, we have also examined cosmographic distance ratio and age crisis revisit by very old cosmological objects at different redshifts.  According to cosmographic distance ratio we found that, higher value of viscosity leads to better agreement with current observational data. Therefore viscous dark energy model constrained  by {\it Planck} TT observation ($\gamma=0.32^{+0.31}_{-0.26}$) is more compatible with data given in \cite{Miyatake:2016gdc}.  In the context of viscous dark energy model, current data for $r$ are mostly affected by CMB used in determining gravitational lensing shear (see Fig. \ref{fig:cgdr}).

Since, there is a competition between Phantom and Quintessence
behavior of viscous dark energy model, it is believed that challenge
in accommodation of cosmological old object can be revisited. We
used 32 objects located in $0.11\le z\le 1.84$ accompanying 9
extremely old globular clusters hosted by M31 galaxy. As reported in
Tab. \ref{tabagetau}, almost tension in age of all old globular
clusters in our model have been resolved at $2\sigma$ confidence
interval. However the value of $\tau$ for  quasar APM $08279+5255$
at $z=3.91$,  with  $t=2.1_{-0.1}^{+0.9}$Gyr is less than unity, but
at $68\%$ confidence limit is accommodated by viscous dark energy
model for all observational catalogues (see Fig.
\ref{fig:age-galaxy}).

For concluding remark we must point out that: theoretical and phenomenological modeling of   dark energy in order to diminish the ambiguity about this kind of energy content have been considerably of particular interest. In principle to give a robust approach to examine dark energy models beyond cosmological constant, background dynamics and perturbations should be considered. The main part of present study was devoted to background evolution. Taking into account higher order of perturbations  to constitute robust approach in the context of large scale structure resulting in more clear view about the nature of dark energy \cite{Amendola:2007rr,Sapone:2012nh,Sapone:2013wda}. The contribution of coupling between dark sectors in the presence of viscosity for dark energy according to our approach is another useful aspect. Indeed considering dynamical nature for dark energy constituent of our Universe is potentially able to resolve tensions in observations \cite{Joudaki:2016kym}.  
These parts of research are in progress and we will be addressing them later.

\begin{acknowledgements}
We are really grateful to Martin Kunz  and Luca Amendola for interesting discussions, during their visiting in modified gravity conference held at Tehran, IRAN. Also we thanks  Nima Khosravi for his comments on framework of this paper. SMSM  was partially supported by a grant of deputy for researches and technology of Shahid Beheshti University. 

\end{acknowledgements}

\appendix

\bibliography{refrences}

\end{document}